\documentclass[12pt,preprint]{aastex}
\usepackage{epsfig}
\usepackage{lscape}
\usepackage{rotating}


\def\ls{{_<\atop^{\sim}}}
\def\gs{{_>\atop^{\sim}}}

\lefthead{Nicastro et al.}

\begin{document}

\title{XMM-{\em Newton} and FUSE Tentative Evidence for a WHIM filament along the Line of Sight to PKS~0558-504} 

\author{F. Nicastro$^{1,2,3}$, Y. Krongold$^4$, D. Fields$^5$, M.L. Conciatore$^3$, L. Zappacosta$^3$, M. Elvis$^3$, S. Mathur$^6$, I. Papadakis$^7$}

\altaffiltext{1}{Osservatorio Astronomico di Roma - INAF, Via di Frascati 
33, 00040, Monte Porzio Catone, RM, Italy}
\altaffiltext{2}{IESL, Foundation for Research and Technology, 711 10, Heraklion, Crete (Greece)}
\altaffiltext{3}{Harvard-Smithsonian Center for Astrophysics, 60 Garden st., Cambridge, MA 02138, USA}
\altaffiltext{4}{Instituto de Astronomia, Universidad Nacional Autonoma de Mexico, Mexico City (Mexico)}
\altaffiltext{5}{L.A. Pierce College, 6201 Winnetka Ave., Woodland Hills, CA, USA}
\altaffiltext{6}{Ohio-State University, Columbus, OH, USA}
\altaffiltext{7}{Physics Department, University of Crete, P.O. Box 2208, GR-710 03 Heraklion, Crete (Greece)}


\begin{abstract}
We present a possible OVIII X-ray absorption line at $z=0.117 \pm 0.001$ which, if confirmed, will be the first one associated with a broad HI Ly$\beta$ (BLB: FWHM=$160^{+50}_{-30}$ km s$^{-1}$) absorber. 
The absorber lies along the line of sight to the nearby ($z=0.1372$) Seyfert 1 galaxy PKS~0558-504, consistent with being a WHIM filament. 
The X-ray absorber is marginally detected in two independent XMM-Newton spectra of PKS~0558-504, a long 
$\sim 600$ ks Guest-Observer observation and a shorter, $\sim 300$ ks total, calibration observation, with a combined single 
line statistical significance of 2.8$\sigma$ (2.7$\sigma$ and 1.2$\sigma$ in the two spectra, respectively). 
When fitted with our self-consistent hybrid-photoionization WHIM models, the combined XMM-{\em Newton} spectrum 
is consistent with the presence of OVIII K$\alpha$ at $z=(0.117 \pm 0.001)$. 
This model gives best fitting temperature and equivalent H column density of the absorber of log$T=6.56_{-0.17}^{+0.19}$ K, 
and logN$_H=(21.5 \pm 0.3) (Z/Z_{0.01\odot})^{-1}$ cm$^{-2}$, and predicts the marginal contribution of only two more lines within the 
XMM-{\em Newton} RGS band pass, NeIX K$\alpha$ ($\lambda=13.45$ \AA) and FeXVII L ($\lambda=15.02$ \AA), both with equivalent 
widths well within the 1$\sigma$ sensitivity of the combined XMM-{\em Newton} spectrum of PKS~0558-504 (EW$^{1\sigma}<3$ m\AA). 
The lack of detection of associated OVI in the archival FUSE spectrum of PKS~0558-504, allows us to infer a tighter 
lower limit on the temperature, of log$T>6.52$ K (at 1$\sigma$). 

The statistical sigificance of this single X-ray detection is increased by the detection of broad and 
complex HI Ly$\beta$ absorption in archival FUSE spectra of PKS~0558-504, 
at redshifts $z=0.1183 \pm 0.0001$ consistent with the best-fitting redshift of the X-ray absorber. 
The FUSE spectrum shows a broad (FWHM$=160^{+50}_{-30}$ km s$^{-1}$)  absorption complex, which we identify as HI Ly$\beta$ 
$z_{BLB}=(0.1183 \pm 0.0001)$. The single line statistical significance of this line is 4.1$\sigma$ (3.7$\sigma$ if systematics 
are considered). 
A possible HI Ly$\alpha$ is marginally hinted in an archival low-resolution ($\Delta\lambda \sim 6$ \AA) IUE 
spectrum of PKS~0558-504, at a redshift of $z=(0.119 \pm 0.001)$ and with single line significance of $1.7\sigma$. 
Thus, the combined significance of the three (XMM-{\em Newton}, FUSE, and IUE) independent tenative detections, is 5.2$\sigma$ 
(5.0$\sigma$ if the HI Ly$\alpha$ is not considered, and 4.6$\sigma$ if the systematics in FUSE are considered). 

The detection of both metal and H lines at a consistent redshift, in this hot absorbing system, allows us to speculate on its metallicity. 
By associating the bulk of the X-ray absorber with the BLB line detected in the FUSE spectrum at $z_{BLB}=0.1183 \pm 0.0001$, 
we obtain a metallicity of 1-4\% Solar. 

Although the absorber is only blueshifted by $\sim -6000$ km s$^{-1}$ from the systemic redshift of 
PKS~0558-504, the identification of the absorbing gas with a high velocity nuclear ionized outflow, is unlikely. The physical, 
chemical and dynamical properties of the detected absorber are all quite different from those typically found in the 
Warm Absorber (WA) outflows, commonly detected in Seyferts and higher luminosity quasars. WA outflow velocities typically span 
a range of few hundreds to $\sim 1-2$ thousands km s$^{-1}$; WA metallicities, when measured, are typically found to be at least 
Solar; high-ionization WAs are virtually always found to co-exist with lower-ionization X-ray and UV phases. 
All this strongly suggests that the absorber, if confirmed, is an intervening WHIM system. 
\end{abstract}



\section{Introduction}
In the present epoch (z$<1-2$), over half (54 \%, Fukugita, 2003) of the baryons are missing. 
They are predicted by Big-Bang Nucleosynthesis (e.g. Kirkman et al., 2003), inferred by density fluctuations of the Cosmic Microwawe 
Background (e.g. Bennet et al., 2003; Spergel et al., 2007), and seen at $z\sim 3$ in the ``Ly$\alpha$ Forest'' (e.g. Rauch, 1998; Weinberg et al., 1997), but by $z < 2$ they are unaccounted for in detected stars and gas (Fukugita, 2003). 
According to hydrodynamical simulations for the formation of structures in a $\Lambda$-CDM 
Universe (e.g. Cen \& Ostriker, 2006), most, if not all, of these missing baryons should 
be in a barely visible ``warm-hot'' phase of the filamentary intergalactic medium (the WHIM). 
The WHIM was shock-heated to temperatures of $10^5-10^7$ K during the continued process of collapse 
and structure formation, and enriched up to Z$^{WHIM}=0.1-1$ Z$_{\odot}$ by galaxy super-winds (GSW, Cen \& Ostriker, 2006). 
These GSWs efficiently removed metals from the cool and dense ISM phase in galaxies and spread them into the tenuous 
($n_e\sim 10^{-6}-10^{-4}$ cm$^{-3}$, i.e. overdensities $\delta \sim 5-500$ compared to the average 
density of the Universe) WHIM phase, enriching the gas to few percent, and up to tens of percent, of 
the solar metallicity values  (e.g. Cen \& Ostriker, 2006). 
WHIM filaments are too tenuous to be detected through their bremsstrahlung and line emission with current X-ray 
instruments (e.g. Yoshikawa et al., 2003). 
However, integrated column densities of high-ionization metal ions along a random section of one of these filaments 
could reach values as high as $\sim 10^{16}$ cm$^{-2}$ (OVII in the soft X-Rays) and 
$\sim 10^{14}$ cm$^{-2}$ (OVI in the Far-Ultraviolet: FUV), imprinting metal absorption lines in the FUV and soft X-ray 
spectra of background sources with equivalent width ranging between EW $\sim 1-20$ m\AA\ (the OVII K$\alpha$ in the X-rays) and   
EW$\sim 10-100$ m\AA\ (the $\lambda\simeq 1032$ \AA\ transition, inthe FUV). 

The most intense of these absorption lines are the OVI$_{1s^22s\rightarrow 1s^22p}$ doublet in the UV 
($\lambda=1031.926, 1037.617$ \AA), and the CV K$\alpha (r)$ ($\lambda=40.268$ \AA), OVII K$\alpha (r)$ 
($\lambda = 21.602$ \AA), CVI Ly$\alpha$ ($\lambda=33.74$ \AA) and OVIII Ly$\alpha$ ($\lambda=18.97$ \AA),  
in the X-ray band. Which of these lines dominates, depends on the ionization state of the gas, that is mainly on its temperature 
(and at a second order on the gas volume density). Dav\'e and collaborators (2001) showed that the 
baryon temperature distribution in the intergalactic space, peaks at log$T\sim 6.6$ K, and is strongly 
skewed toward low temperatures: 50\% of the baryon in the WHIM are found between log$T=5.6$ K and log$T=6.7$ K 
at a weighted average temperature of log$T=5.9$ K, while the low (log$T=(5-5.6)$ K) and high (logT$=(6.6-7)$ K) 
temperature tails of the WHIM temperature distribution, contain 27\% and 23\% of the gaseous 
baryons in the Universe, respectively, around weighted average teperatures of log$T=5.4$ K and log$T=6.7$ 
(Conciatore et al., in preparation). 
So, while the vast majority (73\%) of the baryons in the WHIM should absorb and emit in the X-rays at the wavelengths 
of the He-like (50\%) and/or H-like (23\%) ions of C and O, a substantial fraction of the WHIM (27\%) should be 
detectable in the FUV,  through Li-like C and O transitions. 

These theoretical predictions have been confirmed by FUV observations (e.g. Danforth \& Shull, 2005, 2008; 
Tripp et al., 2008; Richther, 2006). 
Danforth \& Shull (2005) analyzed FUSE data of 31 AGNs with $z<0.15$ to search for OVI absorption counterparts to 129 known 
Ly$\alpha$ absorbers. They found 40 such systems, deriving a $dN_{OVI}/dz(\ge EW_{Thresh})$ 
that agrees strikingly with the latest predictions by Cen \& Fang (2006; see their Fig. 2). 
However, as pointed out recently by Tripp et al. (2008), the vast majority of the associated HI Ly$\alpha$ 
absorption lines are too narrow to be produced in gas with typical WHIM temperatures. 
Danforth \& Shull (2008) proposed that low-ionization metals and narrow HI Ly$\alpha$ trace mildly photoionized medium 
(the low-$z$ Ly$\alpha$ Forest: $\sim 30$\% of the baryons), while narrow-HI/OVI/NV associations 
trace a multiphase intergalactic medium, with a photoionized portion of the intervening filament 
imprinting the narrow-HI absorption and a shock-heated part (WHIM) imprinting the OVI/NV on the UV 
spectra ($\sim 10$\% of the baryons). 
Danforth \& Shull (2008) estimate $\Omega_{WHIM}^{OVI} = 0.34$ \% (down to logN$_{OVI} > 13.4$ cm$^{-2}$). This is $\sim 15$ \% of the ``missing mass'' (strictly speaking, only a lower limit, given the large uncertainties in the ionization correction for the OVI-bearing gas), in good agreement with the WHIM baryon fraction predicted to reside in the low-temperature tail of the WHIM mass-temperature distribution. 

Detecting the bulk of the 'Missing Baryons' in the ``X-Ray Forest'', instead, has proven to be 
extremely difficult. This is because of the unfortunate combination of (a) the still limited 
resolution ($R\sim 400$ at $\lambda = 21.6$ \AA) and the low throughput ($A_{Eff} \sim 20-40$ cm$^2$) of 
the current high-resolution X-ray spectrometers [the XMM-{\em Newton} Reflection Grating Spectrometer (RGS, 
den Herder, et al. 2001) and the {\em Chandra} Low Energy Transmission Grating (LETG: Brinkman et al., 2000)]; 
(b) the lack of bright ($f_{0.5-2 keV} \ge 10^{-11}$ erg s$^{-1}$ cm$^{-2}$) extragalactic point-like 
targets (e.g. Conciatore et al., in preparation); and, (c) the dramatic steepening ($\Delta\alpha \gs 1.5$) of 
the predicted number density of metal WHIM filaments per unit redshift at ion column densities $\gs 10^{15}$ cm$^{-2}$. 

Current evidence is still limited, and highly controversial: in 2005, Nicastro et al. (2005a,b) reported 
the first two $\ge 3\sigma$ detections of absorption systems identifiable with WHIM filaments, 
at $z=0.011$ and $z=0.027$, towards the blazar Mkn~421, which was in outburst. 
One of these two systems is at a redshift consistent with that of a known intervening HI Ly$\alpha$ 
absorber (Shull, Stocke \& Penton, 1996) that line, however, is too narrow, and so the absorbing gas too cold, to be 
physically associated with the X-ray metal filament. 
This claim has been questioned in two subsequent papers by Kaastra et al. (2006) and 
Rasmussen et al. (2007), based on: (a) apparent wavelength inconsistencies for some of the lines, which called into 
question their association with a given WHIM system; (b) statistical arguments for a significantly lower detection significance 
for the WHIM identifications than originally reported by Nicastro et al. (2005a,b); and (c) the lack of detection of these two systems in 
XMM-{\em Newton} RGS calibration spectra of Mkn~421 (but see also Nicastro, Mathur \& Elvis, 2008, for rebuttal arguments). 
Fang, Canizares \& Yao (2007) proposed the detection of an OVIII WHIM filament along the line of sight to the blazar 
PKS~2155-304, once again near the redshift of a known narrow HI Ly$\alpha$ absorber not due to the same gas as 
the putative X-ray filament, because of the line width. 
The above reported pieces of evidence for the existence of an ``X-ray Forest'' at low redshift, have been 
gathered by exploiting the technique of observing random lines of sight toward the brightest possible X-ray sources, 
with exposures long enough to reach extremely high S/N spectra. This technique has the advantage of not being biased toward 
the strongest WHIM systems, and so allows probes of the bulk of the WHIM mass distribution. 
However, the WHIM is predicted, and in the FUV has most likely been found (e.g. Stocke et al., 2006b), to correlate strongly 
with galaxy concentrations in the Universe: the densest of such concentrations are also supposed to host the nodes, and so 
the densest and hottest parts, of the WHIM network. 

An alternative technique, therefore, is to target the WHIM search in the 
X-rays along those lines of sight toward bright background X-ray sources that intercept large scale structure concentrations. 
By exploiting this technique Buote et al. (2009), reported a $\sim 3\sigma$ detection of a high column density (N$_{OVII} \ge 10^{16}$ 
cm$^{-2}$) OVII K$\alpha$ absorption line, along the line of sight to the blazar H~2356-309, at the redshift of a very large 
concentration of galaxies in the Sculptor Wall. This technique has the disadvantage of probing 
only the extreme high-temperature/density and low-mass fraction ends ($\sim 23$\%) of the WHIM distribution. 

All the above proposed WHIM detections so far, suffer the serious handicap of lacking the detection of a clearly associated 
HI counterpart. 
The detection of broad HI Lyman lines (mostly Ly$\alpha$ at $\lambda=1215.67$ \AA\ - BLA - and 
Ly$\beta$ at $\lambda=1025.72$ \AA\ - BLB) is vital for a proper assessment of the WHIM metallicity and mass, 
and require the use of both X-ray and FUV facilities. 
Without HI, metallicity cannot be determined and so a cosmological mass density $\Omega_b^{WHIM}$ cannot be derived. 

\noindent
The expected thermal FWHM of an absortion line imprinted from a gas of particles of mass $m$ at the 
equilibrium temperature $T$, is given by $2\sqrt{2ln2}$ times the variance of the velocity distribution (the Doppler parameter $b$), 
that is FWHM$=2\sqrt{2ln2} \sqrt{kT/m}$. For HI, this implies a broadening of FWHM$\sim 380$ km s$^{-1}$ at log$T=6.5$
\footnote{For heavier elements, the width scales with the inverse of the root square of the atomic weight. So, for example, for 
O, Ne and Fe, at log$T=6.5$, FWHM=95, 85 and 51 km s$^{-1}$.}
. 

\noindent
The HI fraction, relative to the total H, in gas with temperatures in the log$T=(5.8-6.3)$ range (where 50\% of the WHIM should be, 
e.g. Dav\'e et al., 2001), spans the interval $10^{-6.2}-10^{-7.2}$. Thus only very shallow (EW(Ly$\alpha \simeq 8-80$ m\AA) and 
broad HI absorption lines are expected to be imprinted onto FUV spectra by WHIM filaments, requiring spectra with  
S/N$\simeq 10-100$ per resolution element to make them detectable. 

Here we present the first tentative detection of a hot (X-ray) WHIM filament along the line of sight to the $z=0.1372$ Seyfert galaxy 
PKS~0558-504, that is physically associated with a BLA HI absorber. 
A full band analysis of the XMM-{\em Newton} RGS spectrum of PKS~0558-504, is deferred to a companion 
paper (Papadakis et al., 2009, submitted). 
In \S 2 we present the XMM-{\em Newton}, FUSE and IUE data that we use in our analysis. \S 3 and 4 are dedicated to 
the X-ray and FUV data reduction and analysis. In \S 5 we critically discuss our findings, and in \S 6 we summarize our conclusions. 

\section{XMM-{\em Newton}, FUSE and IUE Data of PKS~0558-504}
Between February 2000 and October 2001, PKS~0558-504 was the subject of an extensive calibration campaign with XMM-{\em Newton}, 
with a total expsoure of 312 ksec (hereinafter 'CAL spectrum'). 
More recently, in September 2008, PKS~0558-504 was re-observed by the XMM-{\em Newton} for 618 ksec, as part of an approved 
cycle 8 Guest Observer program (PI. S. Papdakis; hereinafter 'GO spectrum'). 
In this work we make use of the RGS spectra extracted from both data sets. We focus mostly on the analysis of the longer and much 
higher S/N 2008 (16 vs 9, at 25 \AA) GO spectrum. We use the lower S/N 2000-2001 CAL spectrum for consistency checks. 

PKS~0558-504 was observed by the FUSE satellite twice: on 1999, December 10 and 2001, November 7, with net exposure times
of 47.2 ksec and 48.3 ksec, respectively. In our analysis we co-add the two spectra from the Lif2A channel (with the highest efficiency in the 1100-1200 \AA\ interval) to maximize the S/N and perform our search for intervening BLAs and BLBs in this co-added FUSE spectrum
\footnote{We also reduced and analyzed the sum of the two spectra from the Lif1B channel, which is sensitive (though with about half of the effective area of the Lif2A channel) in a wavelength interval similar to that of the Lif2A. However, the Lif1B channel is unreliable at $\gs 1140$ \AA\ 
(The FUSE Data Handbook, Chapter 7, \S 7.3.2, http://archive.stsci.edu/fuse/DH\_Final/Factors\_Affecting\_FUSE\_Data.html), which is 
the region of interest for our analysis. Therefore we decided to consider only the Lif2A channel in our analysis (but see also footnote 9, \S 4).}
.

Finally, PKS~0558-504 has been observed three times with the 'Short-Wavelength' spectrometer of the IUE satellites (the only one covering the 
wavelength range of interest), on 1987, September 22, and 1989, November 14-15. In our analysis we use the highest S/N of the three spectra 
(with a net exposure time of 16.2 ksec), the one taken on 1989, November 15, and check the other two spectra for consistency. 

Table 1 lists the journal of the XMM-{\em Newton}, FUSE and IUE observations used in this work. 
 For all our spectral fitting we use the fitting package {\em Sherpa}, of the Chandra Interactive Analysis of Observation ({\em Ciao}) software (v. 4.1.2; Freeman, Doe \& Siemiginowska, 2001). 

\section{The RGS Spectrum of PKS~0558-504}

\subsection{RGS Data Reduction and Analysis}
The 2008 GO RGS data of PKS~0558-504 were reduced as in Papadakis et al. (2009), and the same standard reduction procedure was applied to the 2000-2001 calibration observations of this target. The data were cleaned for periods of high background activity, by excluding events taken within time-intervals where the background deviated positively by its average level by more than $2\sigma$. 
Finally RGS1 and RGS2 spectra and responses of the single observations of the two datasets (CAL and GO) were co-added and averaged by using the {\em rgscombine} tool. 
This produced total net exposures of 480 ksec (RGS1) and 466 ksec (RGS2), for the GO dataset, and 309 ksec (RGS1) and 283 ksec (RGS2) for the CAL  dataset. 
Despite a ~30\% higher average flux level of PKS~0558-504 during the CAL observations, compared to the GO observations, and a relatively small difference of only 65\%  in the net exposure times of the two datasets, the S/N of the GO spectra is significantly (factor of $\sim 2$) higher than that of the CAL spectra. In half RGS resolution element (30 m\AA), at 0.5 keV the co-added RGS1 CAL spectrum has S/N=9, compared with S/N=16 for the GO spectrum. This difference in S/N is due to the much higher (more than an order of magnitude) level of the average background during all 
individual CAL observations, compared with the GO observations. 
This is clearly seen in Figure 1, where we plot the combined GO RGS1 20-24 \AA\ background-subtracted source (green filled circles and errorbars) and background (blue filled circles and errorbars) spectra of PKS~0558-504, together with the combined CAL RGS1 20-24 \AA\ background-subtracted source (black empty squares and errorbars) and background (red empty squares and errorbars) spectra of PKS~0558-504. 
We therefore decided not to further combine the RGS spectra of the two datasets, and to analyze them separately and simultaneously. 

In our spectral analysis we bin the RGS spectra two 45 m\AA\ bin (about half the nominal FWHM resolution of the RGSs). 

\subsection{X-Ray Spectral Fitting}
In our companion paper (Papadakis et al., 2009), we present the broad-band 0.4-2 keV spectral analysis of the 2008 RGS spectra of PKS~0558-504, 
and show that the soft X-ray continuum of PKS~0558-504 is well modeled by a broken power law with a break energy of E$_{brk}\sim 1.5$ keV, 
attenuated by Galactic ISM absorption. More importantly we exclude the presence of a Warm Absorber with typical column densities and ionization 
parameters (Papdakis et al., 2009). This is consistent with the absence of associated, intrinsic, Narrow Absorption Lines (NALs) in the Far Ultraviolet spectrum of PKS~0558-504 (Dunn et al., 2007). 
However, Papadakis et al. (2009) also note that weak absorption line-like features are detected in the spectrum of PKS~0558-504, in the narrow 20-24 \AA\ spectral interval. 
These three lines, and in particular the line at $\lambda=21.18$ \AA\, are the subject of this study. 

We first explored the 20-24 \AA\ region of the GO and CAL RGS1 spectra of PKS~0558-504 (the RGS2 is blind in this wavelength range, due to the lack of a read-out CCD chip). We fitted the data simultaneously, with two local continuum models, each including a power law attenuated by 
neutral absorption with the metallicity set to Anders \& Grevesse (1989). While such a model gives a statistically acceptable fit for the GO 
spectrum of PKS~0558-504, it leaves broad and systematic residuals in the CAL spectrum. This is because the {\em rgscombine}-averaged CAL response matrix fails to properly reproduce the depth of the several effective area features present in this portion of the RGS1 spectrum. 
We therefore added a number of FWHM$\sim 0.5-1$ \AA\ broad emission and absorption gaussians to the best fitting continuum model 
of the CAL spectrum of PKS~0558-504, until we reached a statistically acceptable fit. 
Three narrow regions with a deficit of counts are left in the GO and CAL residuals to the best fitting continuum models, at $\lambda \sim 23.5$ \AA, 
$\lambda \sim 21.6$ \AA\ and $\lambda \sim 21.2$ \AA.  
To model these residuals we added three Gaussians to our best-fitting continuum models and refitted the data leaving all Gaussians parameters free to vary independently in the fit to the GO spectrum, but linking the line energies and Gaussian widths in fitting the CAL spectrum to those of the corresponding parameters of the fitting model of the GO spectrum. 
The first three rows of Table 2 show the best fitting parameters of these three absorption lines. Errors on wavelengths and 
redshifts are assumed to be equal to 1 bin-size, 30 m\AA, if unresolved by the fitting routine). 
Figure 2 shows the 20-24 \AA\ RGS1 CAL and GO data of PKS~0558-504, with their best-fitting models  superimposed.  
The first two of these lines, at $\lambda=23.52 \pm 0.03$ \AA\ and $\lambda=21.61 \pm 0.03$ \AA, are easily identified with the K$\alpha$ inner shell transition from atomic OI in the Interstellar Medium of the Galaxy, and the K$\alpha$ He-like transition from OVII either in our Galaxy halo or in the Local group (e.g. Bregman et al., 2007 and references therein). 
The third line, at $\lambda=21.17 \pm 0.04$ \AA, appears marginally resolved (but still consistent with an unresolved Gaussian at a 1$\sigma$ confidence level
\footnote{the {\em projection} routine in Sherpa, which we use here to compute uncertainties and upper limits, finds only an 
upper limit for the line FWHM, at a 1$\sigma$ confidence level. This is because of the high non-gaussianeity of the RGS Line 
Spread Function (e.g. Williams et al., 2005), which makes the effective LSF FWHM significantly broader than the nominal Gaussian 
equivalent LSF FWHM of 60 m\AA (corresponding to $\sim 900$ km s$^{-1}$ at 21.2 \AA).}
) in the RGS1 (FWHM$\ls 1200$ km s$^{-1}$: Table 2) and has no obvious identification with transitions 
at $z\simeq 0$. 
If red-shifted to the systemic redshift of PKS~0558-504 ($z=0.1372$, REFs), this line would have a rest-frame wavelength of $\lambda=18.62$ \AA, close to the rest-frame wavelength of the OVII K$\beta$ transition. However, in this case, a much stronger (factor $\sim 6.4$) OVII K$\alpha$ line 
\footnote{together with several other strong lines from He- and H-like ions of C, N and Ne}
with EW$\sim 60$ m\AA\ should be visible at $\lambda= 24.55$ \AA. 
Such a line is not seen in the data (see also Papadakis et al., 2009) at a 3$\sigma$ limit of EW$< 18$ m\AA
\footnote{we also caution, however, that although the RGS1 spectra of PKS~0558-504 are free of instrumental feature at, exactly, 24.55 \AA, this spectral region of the RGS1 contains several instrumental features, which might reduce, due to calibration uncertainties, the true sensitivity of the RGS1 spectrum at 24.55 \AA. See also \S 5.2}
. Moreover, the absence of intrinsic OVI absoprtion in the FUSE data of PKS~0558-504 (Dunn et al., 2007), makes this interpretation less likely. 

The next possible identification is with an intervening absorber, between the redshift of PKS~0558-504 and us. This imposes the condition 
that the line has a rest-frame wavelength in the 18.6-21.2 \AA\ spectral interval. 
The strongest resonant absorption line expected in this spectral range is the OVIII Ly$\alpha$ transition, with an unresolved doublet at 
an oscillator strength weighted average wavelength $<\lambda>=18.97$ \AA. This would give a redshift of $z=0.116 \pm 0.002$ for the intervening 
absorber. 

\subsection{Consistency Test with a Self-Consistent WHIM Spectral Model}
To check whether the presence of an OVIII$_{K\alpha}$ WHIM filament at $z=0.116 \pm 0.002$ along the line of sight to 
PKS~0558-504, is consistent with the full band XMM-{\em Newton} RGS spectra of this target, we used our hybrid ionization 
(collisional + photoionization by the meta-galactic UV and X-ray background at a given redshift) WHIM spectral 
models (an evolution of our PHotoinized Absorber Spectral Engine - ``PHASE'' -, Krongold et al., 2003), to simultaneously fit the 
RGS1 and RGS2 Cal and GO data of PKS~0558-504. The fitting model includes our best-fitting continuum model (a combination of 
power-laws and broad gaussians to cure residual calibration uncertainties), two negative gaussians to model the OI$_{K\alpha}$ and OVII$_{K\alpha}$ 
absorptions at $z\simeq 0$, and a WHIM model. In the X-ray band alone one cannot constrain the absolute metallicity of the absorber 
(for which detection of HI is needed), since the ionization balance of an hybridly ionized cloud of gas (i.e. the fractional ion abundances), parameterized by the temperature of the absorber, at equilibrium, is virtually independent on metallicity (see also \S 5.2). 
Thus, for fitting purposes only, we froze the WHIM metallicity to Solar. 
The turbulent velocity of the absorber was frozen to 400 km s$^{-1}$ (about 1/2 of the RGS FWHM resolution at 22 \AA). 
In a hybrid-ionization WHIM model, photoionization by the meta-galactic UV and X-ray photon field, is only a secondary ionization mechanism, 
and its realtive importance, compared to the collisional shock mechanism, depends only on the volume density of the WHIM filament: 
the lower the density the more important the photoionization contribution. In such models the absorber density is thus highly degenerate with the absorber temperature and equivalent H column density, and so is difficult to constrain. 
We therefore froze the density of the absorbing gas to a typical expected WHIM value of n$_b=10^{-5}$ cm$^{-3}$ (overdensity $\delta\simeq 50$ compared to the average baryon density in the Universe), corresponding to an ionization parameter (the ratio between the photon density 
at the surface of the absorbing cloud and the baryon density in the cloud) of log$U=-1.9$ at $z=0.1$. At such densities (and higher), the photoionization contribution is essentially negligible, and the model reduces to a pure collisional ionization model. 
The remaining WHIM parameters, namely the gas temperature and equivalent hydrogen column density, and the redshift of the WHIM 
filament, were left free to vary in the fit. 
The best-fitting WHIM model (Fig. 3, black curve) has log$T=6.56_{-0.17}^{+0.19}$, logN$_H=19.5 \pm 0.3$ (for $Z=Z_{\odot}$) and 
$z=0.117 \pm 0.001$. 
This model fits the most prominent OVIII K$\alpha$ line at $z=0.117$, and predicts only two additional weak absorption lines ($EW\ls 1$ mA): the NeX K$\alpha$, at $\lambda = 15.02$ \AA\ (WHIM filament frame) and a strong (oscillator strength 2.95) FeXVII L line at $\lambda = 16.77$ \AA (Fig. 3, black curve). 

Figure 4, 5 and 6 shows the two portions of the RGS CAL and GO spectra of PKS~0558-504, in which the lines predicted by the best-fitting WHIM model lie. The superimposed red curve is the best fitting WHIM model convolved with the instrumental responses. 
The OVIII K$\alpha$ line is detected both in the CAL and GO spectra (Fig. 4). The position of the predicted NeIX K$\alpha$ and FeXVII L lines is marked in Fig. 5 (RGS1 and RGS2 GO spectra of PKS~0558-504) and Fig. 6 (RGS1 ands RGS2 CAL spectra of PKS~0558-504): the strength of these lines 
is $< 1/3$ the minimum detectable 1$\sigma$ EW in the combined CAL+GO spectrum (Table 2). 
We note the hint of the presence of the FeXVII L line at 16.77 \AA\ in both the RGS1 GO (Fig. 5, top panel) and RGS2 CAL (Fig. 6, bottom panel) spectra of PKS~0558-504. 

In Figure 3 we plot the best fitting WHIM model (central, black curve) together with its negative (lower, red curve) and positive (upper, blue curve) 1$\sigma$ temperatures. Both lower and upper limits on the temperature are weak. The strongest X-ray lower limit diagnostics in the model is the increasing strength of OVII K series absorption lines. However, the sensitivity of the RGS1 GO and CAL spectra of PKS~0558-504 at the wavelength of the strongest line of this series (the OVII K$\alpha$ line) is dramatically reduced by the presence of 
an instrumental feature due to a bad pixel in the read out dispersion detectors. On the other end, at higher temperature, the limit is set by the decrease of opacity at the wavelengths of the OVIII K$\alpha$ transitions, which is only detected at a combined single-line significance level of $2.8\sigma$ in the data. 
This can be also seen in Figure 7, where we show the 68\%, 90\% and 95\% log$T$ - logN$_H$ contours of the best-fitting WHIM model. The parameters are clearly correlated and the temperature poorly constrained both at low and high temperatures. However, low temperatures and high column densities are excluded at high confidence levels. 

\section{The FUSE and IUE Spectra of PKS~0558-504}

We searched the The Multimission Archive at STscI (MAST) archive
\footnote{http://archive.stsci.edu/index.html}
for FAR-UV data of PKS 0558-504 and found several data sets, taken with FUSE, HST and IUE satellites (Table 1). Our main goal was to search for 
HI (Ly$\alpha$ and Ly$\beta$) and OVI absorption either at or near the systemic redshift of PKS~0558-504, or at a redshift consistent with that of the putative X-ray WHIM filament, tentatively detected in the RGS spectrum of PKS~0558-504. 
The available archival HST spectrum of PKS~0558-504 was taken with the STIS-G230 gratings, and so covers a wavelength interval ($\lambda=
2760-2910$ \AA) not relevant to this analysis. The FUSE and IUE spectra, instead, cover two important regions: (a) the HI Ly$\beta$ and OVI regions (FUSE) and (b) the HI Ly$\alpha$ region (IUE). We retrieved these data sets from the MAST archive, and analyzed them to search for a secure identification of the X-ray absorber. 

FUSE observed PKS~0558-504 twice, in two different programs (P101, Sembach 1999, and C149, Prochaska 2002).  All FUSE observations after the end of operations were re-reduced using the final version of the CalFUSE pipeline  (v3.2.3, Dixon et al., 2007), and we used the resultant science calibrated files.  
We used specific IRAF scripts 
\footnote{http://fuse.pha.jhu.edu/analysis/IRAF\_scripts.html}
to co-add the data from the Lif 2A channel, which covers the wavelength range 1087-1181 \AA, with a total exposure time of 90.3 ks. 
The spectrum was then binned to two different resolutions: the natural resolution of FUSE (3 bins; 11 km/s) and a poorer resolution (20 bins; 70 km/sec) with higher S/N per bin, specifically to search for broad, shallow absorption features. 

We use the fitting package {\em Sherpa} in Ciao (v. 4.1.2; Freeman, Doe \& Siemiginowska, 2001), to search the FUSE spectrum for: (a) AGN intrinsic absorption between -1000 and +1000 km s$^{-1}$ of the systemic redshift of PKS~0558-504; (b) intervening absorption at a redshift consistent with that of the putative X-ray WHIM filament.  Specifically, for each of the two possible systems, we searched for absorption lines from: (1) the NIII triplet at $\lambda=989.80$ \AA, $\lambda=991.51$ \AA, and $\lambda=991.57$ \AA; (2) the OVI doublet at $\lambda=1031.93$ \AA\ and $\lambda=1037.62$ \AA; and (3) the HI Ly$\beta$, at $\lambda=1025.72$ \AA. 
We confirm the non-detection of intrinsic (i.e. nuclear) ionized or neutral absorption, down to $EW \ls 30$ m\AA\ (3$\sigma$) of Dunn et al. (2007). 
However, we do detect several absorption lines in the wavelength range 1137-1155 \AA\ (Figure 8). 
The strongest of these lines are easily identified as Galactic FeII, PII and possibly H2 transitions. These lines are all relatively narrow 
, with typical FWHM$\simeq 50-70$ km s$^{-1}$
\footnote{With the exception of the high velocity component of the FeII($\lambda 1144.94$) absorber, at $\lambda = 1145.7$ \AA, for which we measure FWHM$=(100 \pm 40)$. This absorber could be at least partly contaminated by intervening HI Ly$beta$ absorption at $z=(0.1170 \pm 0.0001)$, i.e. still consistent with the X-ray redshift of the putative OVIII intervening absorber, and so possibly 
representing a second HI component associated with this hot intervening WHIM filament.}
, and exhibit two velocity components at $v_1\simeq 0$ and $v_2 \simeq +200$ km s$^{-1}$ (for the low-ionization atomic transitions)
\footnote{There are also two known velocity components in high-ionization OVI Galactic (or Galactic halo) absorption, along the line of sight to PKS 0558-504, in the velocity ranges $-115$ -- +135 km s$^{-1}$ (Savage, 2003) and +210 -- +315 km s$^{-1}$ (Sembach et al., 2003).} 
, and $v_1\simeq 0$ and $v_2 \simeq +70$ km s$^{-1}$ (for the molecular transition: Wakker, 2006): Figure 8. 

However, we also detect a broader absorption complex at the centroid wavelength of $\lambda=1147.1$ \AA, and at single-line statistical significances of 4.1$\sigma$ (Figure 8)
\footnote{The associated 1$\sigma$ error here, and anywhere else in the paper, is only statistical. By taking into account the vagaries of continuum placement and fixed-pattern noise in the Lif2A FUSE spectrum of PKS~0558-504, we estimate an additional 14\% systematic 1$\sigma$ uncertainty. This would reduce the significance of the broad line at  $\lambda=1147.1$ \AA, from 4.1$\sigma$ to $3.7\sigma$.}
. We model this absorber with a negative Gaussian, and list its best-fitting parameters and equivalent width in Table 1
\footnote{We note that we this line is also clearly visible in the co-added Lif1B spectrum, with a slightly larger, but consistent, EW. However, due to the unreliability of the LIf1B channel at $\lambda \gs 1140$ \AA\ (see footnote 1, \S 2), we conservatively decided not to consider this as a supporting evidence.}
. 

This absorption line is broad (FWHM$=160^{+50}_{-30}$ km s$^{-1}$)  and cannot be identified with any strong Galactic transition. 
We identify this absorber as a broad HI Ly$\beta$ (BLB) at $z_{BLB}=(0.1183 \pm 0.0001)$. 
The X-ray redshift of the OVIII K$\alpha$ absorber is consistent with $z_{BLB}$. 
At the best fitting X-ray temperature of logT=6.56 K, the thermal speed of protons is $\sqrt{2} b=250$ km s$^{-1}$ and so the 
expected broadening of HI absorption lines is FWHM$\simeq 400$ km s$^{-1}$. 
The width of the $z_{BLB}$ HI absorber is marginally consistent, at a 2.5$\sigma$ level, with the expected broadening (Fig. 9).  
Here we therefore assume that the HI BLB is associated with the bulk of the X-ray OVIII absorber. 
 
There is no detectable OVI or NIII absorption associated with the $z_{BLB}$ HI absorber. 
The non-detection of OVI puts a much more stringent lower limit on the temperature of the detected WHIM filament. 
At the best fitting X-ray temperature the relative fraction of OVI, $\xi_{OVI} = 2.0_{1.8}^{+9.0} \times 10^{-4}$, predicts (for unsaturated lines) EW(OVI)$=7.6^{+40.6}_{-6.8}$ m\AA. From the FUSE data we have EW(OVI)$_{z=0.118} \ls 12$ m\AA\, at a 1$\sigma$ confidence level (Table 2). By linearly interpolating between expected OVI fractions and EWs, we derive the 1$\sigma$ upper limit on the fraction of OVI corresponding to the measured EW(OVI) upper limit, and from this the 1$\sigma$ lower limit on the temperature of the absorber. This gives: logT$>6.52$. Thus by combining the UV and X-ray constraints on the temperature of the WHIM filament, one gets logT$=6.56_{-0.04}^{+0.17}$ (Table 3). 

The IUE satellite observed PKS~0558-504 twice with the 'Short Wavelength Primary' (SWP), low-resolution ($\Delta\lambda=6$ \AA) spectrometer, on 1987 September 22, and 1989, November 14-15. We retrieved the data products of these three observations from the MAST archive
\footnote{http://archive.stsci.edu/index.html}
, and analyzed them with {\em Sherpa} (v. 4.1.2
\footnote{http://cxc.harvard.edu/ciao/index.html}
; Freeman, Doe \& Siemiginowska, 2001). 
The three SWP Low-Resolution spectra (LR) of PKS~0558-504, are dominated by the intrinsic nuclear AGN continuum and emission line spectrum. 
However, the two spectra with highest S/N (obsid 37589 and 37604; Table 1) show also marginal evidence for a narrow absorption feature imprinted on the blue wing of the HI Ly$\alpha$ broad emission line of PKS~0558-504, at $\lambda \sim 1360$ \AA\ (Figure 10). 
We fitted the three spectra simultaneously with a model, consisting of a powerlaw plus narrow and broad components of the HI Ly$\alpha$ and NV AGN emission lines and a negative absorption line to model the absorption feature at $\lambda \simeq 1360$ \AA. 
The best-fitting parameters for this absorption line are listed in Table 2. We tentatively identify this low-significance (1.7$\sigma$) absorber 
as an intervening broad HI Ly$\alpha$ (BLA) at $z_{BLA}=0.119 \pm 0.001$, a redshift consistent with that of the $z_{BLB}$ HI absorber detected by FUSE. 
Figure 10 shows a portion of the highest S/N 1989 IUE-SWP LR (obsid 37604) spectrum of PKS~0558-504, with our best-fitting superimposed, 
and the residuals after subtracting the Gaussian at $\lambda=1360 \pm 1$ \AA. 

\subsection{Summary of the Observational Evidence}
The Far-UV and X-ray spectra of PKS~0558-504 show evidence for complex absoprtion, which is unlikely to be intrinsic to the 
AGN itself (\S 5.1). Figure 11 summarizes these conclusions, by showing the residuals (in $\sigma$), in velocity space, to the best fitting continuum models to the FUSE (top panel), IUE-SWP LR (1989; second panel from the top) and XMM-{\em Newton} (GO: 3rd and 5th panels from the top; CAL: 4th panel from the top) spectra of PKS~0558-504. The residuals are centered around the absorption lines that we identify as HI Ly$\beta$, HI Ly$\alpha$, OVIII Ly$\alpha$ and FeXVII L ($\lambda=15.015$ \AA, rest frame), at a mean common redshift of $z=0.118 \pm 0.001$ (all 
lines are consistent with this redshift). The red arrows in the figures indicate the best-fitting centroid of each line in their respective spectra. 

The FeXVII L $\lambda=15.015$ \AA) line (bottom panel) is only seen at 1.5$\sigma$ and 2$\sigma$ in the RGS1 GO and RGS2 CAL spectra (Figs. 5 and 6), and is not visible in either the RGS2 GO or in the RGS1 CAL spectra, although all data are consistent with the presence of this line at the strength predicted by the best-fitting WHIM model. 
The OVIII K$\alpha$ line is present both in the GO (3rd panel from the top) and CAL (4th panel from the top) RGS1 spectra (the RGS2 is blind in this spectral region). 
Finally, possible HI lines are seen both in the FUSE (top panel) and IUE-SWP LR spectra: a BLB is detected in the high resolution FUSE spectra, at redshifts $z_{BLB}=(0.1183 \pm 0.0001)$, and an unresolved BLA is hinted in the IUE spectra at $z_{BLA}=(0.119 \pm 0.001)$. Both redshifts are consistent with that of the X-ray absorbers, within their 1$\sigma$ uncertainties. 
Finally we note that the measured ratio EW$_{BLA}$/EW$_{BLB}$$= (15 \pm 10)$, between the equivalent width of the low-significance 
BLA line (IUE) and the BLB line (FUSE) is consistent, within the large statistical errors, with the predicted value of 
(EW$_{HI}$)$_{Ly\alpha}$/(EW$_{HI}$)$_{Ly\beta}$$\simeq 7.4$, for unsaturated lines
\footnote{For unsaturated lines, the equivalent width ratio of two electronic transitions $1$ and $2$ from the same series of the same ion $X^i$, is linearly related to the product between their oscillator strength ($f$) ratio times the square of their rest-frame wavelength ($\lambda$) ratio: i.e.  (EW$_{X^i}$)$_1$/(EW$_{X^i}$)$_2 \simeq (f_{X^i})_1 / (f_{X^i})_2 \times (\lambda_1 / \lambda_2)^2$. Thus, for the Ly$\alpha$ and Ly$beta$ transitions of HI, one expects: (EW$_{HI}$)$_{Ly\alpha}$/(EW$_{HI}$)$_{Ly\beta}$$\simeq 7.4$.} 
. 

The combined statistical significance of this detection is $5.2\sigma$ (4.6$\sigma$ if systematics are considered, and the 
small conribution from the putative HI BLA is not included). 

\section{Discussion}

\subsection{Ruling Out Intrinsic Absorption}
The Far-UV and X-ray absorber along the line of sight to PKS~0558-504, is only $\sim -6000$ km s$^{-1}$ from the systemic redshift of 
PKS~0558-504. In principle, then, it could be identified with a high-velocity outflow ejected from the Seyfert's nuclear region in the direction of our line of sight. Such phenomena are common in AGNs, and are known with the name of ``Warm Absorbers (WAs, in X-rays: e.g. George et al. 1998;  Piconcelli et al., 2005) or ``Narrow Absorption Line'' absorbers (NALs, in the UV: e.g. Crenshaw, Kraemer \& George, 2003). 
However, we note that, the physical, chemical and dynamical properties of the detected absorber are all quite different from those typically found in WAs and NALs, commonly detected in Seyferts and higher luminosity quasars. 

WA/NAL outflow velocities typically span a range of few hundreds to  few thousand km s${-1}$, with $\gs 80$\% of the objects having $v_{out} \ls 1500$ km s$^{-1}$ (e.g. Blustin et al., 2005 for WAs in X-rays, and Kriss, 2002  Crenshaw, Kraemer \& George, 2003 for NALs in the UV) and are normally found to be present in several physical (temperature, and densities) and velocity components in a single object both in X-rays (e.g. Krongold et al., 2003, 2005, 2009)  and in the UV (e.g. Kriss, 2002; Crenshaw, Kraemer \& George, 2003). 
Here we measure an extreme velocity outflow of $-6000$ km s$^{-1}$ and do not detect any other X-ray or FUV systems at lower velocity. 
The existence of very high ionization (FeXXV and FeXXVI) X-ray outflows, with sub-relativistic (i.e. $\sim 0.1 c$) outflow velocities has been 
recently proposed for few objects (Reeves et al., 2009, and reference therein), but the existence and identification of these features are still highly debated (e.g. Uttley, 2009). In all such proposed cases, however, the high-ionization absorber has very high equivalent hydrogen column density ($few \times 10^{23}$ cm$^{-2}$), is seen often together with other lower velocity, lower ionization components, and in some cases requires super-solar Fe abundance (Reeves et al., 2009 and references therein). We checked the high S/N XMM-{\em Newton} GO EPIC spectrum of PKS~0558-504 for the presence of blue-shifted He- or H-like Fe absorption, and did not find any down to an equivalent hydrogen column densities of $\gs 10^{21}$ cm$^{-2}$. 

WA metallicities, when measured, are typically found to be at least Solar (e.g. Fields et al., 2005, 2007), in sharp contrast with the 1-5\% Solar value we measure for this absorber. 
We stress that our metallicity estimate for the $z=0.118 \pm 0.001$ absorber is virtually model-independent. The HI fractions in purely photoionized or collisionally ionized clouds of gas that best-fit the RGS spectra of PKS~0558-504, are almost indistinguishable from the HI fraction that we 
derive from our best-fitting hybrid-ionization WHIM models. So, the metallicity-correction needed to reconcile the HI BLB (FUSE) and BLA (IUE) column density measurements with the X-ray equivalent hydrogen column density estimate, is independent on the model 
used to fit the X-ray data. 

Finally we stress that high-ionization WAs are virtually always found to co-exist with lower-ionization phases (e.g. HIP and LIP, Krongold et al., 2003, Andrade-Velasquez et al., 2009),  which in turn are often one to one associated with NAL components in the UV. In the case of PKS~0558-504 we do not detect any intrinsic OVI or CIV absorption (see also Dunn et al., 2007). 
All this makes the identification of the detected absorber with a high velocity nuclear ionized outflow, unlikely. 

A WA with unprecedented velocity, metallicity and ionization phases, appears to us to be less likely than the well-predicted WHIM possibility. 
We therefore tentatively identify this complex absorber with an intervening WHIM filament at the mean common redshift of $z=0.118 \pm 0.001$, 
and in the following discussion associate the X-ray absorber with the Far-UV BLB and BLA detected in the FUSE and IUE spectra. 

\subsection{Metallicity and Thickness of the WHIM Filament}
In principle, the detection of clearly associated metal and HI transitions from the same WHIM filament, allows one to measure the 
the metallicity of the absorber. 
In our case, the spectral resolution of the X-ray data is much poorer than that of the superior quality FUSE data, and the best-fitting 
width of the HI BLB absorber tentatively identified at a redshift consistent with that of the X-ray absorber (FWHM=$160_{-30}^{+50}$ 
km s$^{-1}$) , is narrower than the expected HI BLB thermal width in gas with the temperature inferred from teh X-ray data. 
Thus a direct association between the HI BLB and the OVIII absorbers cannot be clearly established, based on these data. 
However, both the redshifts of the HI BLB and the OVIII absorbers, and the thermal width of the HI BLB lines, are consistent, within 
the large uncertainties (e.g. Fig. 9), with the hypothesis that the bulk of the HI and the OVIII absorbers are imprinted by the same 
gas. Under this assumption, we then estimate the metallicity of the putative WHIM filament, and note that this estimate would 
translate in a lower limit if the BLB and the metal absorbers were structured or not one-to-one associated. 

From the measured temperature, logT$=6.56_{-0.04}^{+0.19}$ we derive the relative fraction of HI compared to total H: $\xi_{HI} =   
(5.4^{+0.5}_{-2.0}) \times 10^{-8}$. As already mentioned in /S 3.3, the best-fitting temperature of the absorber is virtually 
independent on metallicity. 
We checked this by building two additional grids of models (other the one with $Z=Z_{\odot}$ used to fit the X-ray data) with metallicity of $Z=0.1 Z_{\odot}$ and $Z=0.01 Z_{\odot}$. Figure 12 shows the OVIII to HI fractional abundance ratio as a function of the temperature of the gas, for $Z=Z_{\odot}$ (black, solid curve), $Z=0.1 Z_{\odot}$ (red, solid curve)and $Z=0.01 Z_{\odot}$ (green, dashed curve). The ionization balance of the gas, at equilibrium, is virtually independent on the value of $Z$. 
From the metal-independent ionization correction value (i.e. $\xi_{HI} =   (5.4^{+0.5}_{-2.0}) \times 10^{-8}$), and the intrinsic (i.e. at z$=0.1183$) HI column of N$_{HI} = (9 \pm 2) \times 10^{13}$ cm$^{-2}$ (infered from the measured EW(BLB)$ = 66 \pm 16$ m\AA\ and FWHM(BLB)$=160^{+50}_{-30}$ km s$^{-1}$), we can thus derive the expected equivalent H column density of the absorber, which turns out to be: N$_H = (1.7_{-0.4}^{+0.7}) \times 10^{21}$ cm$^{-2}$ (Table 3). 
Our best fitting WHIM model measures N$_H = (3.2^{+3.1}_{-1.6}) \times 10^{19} (Z/Z_{\odot})^{-1}$ cm$^{-2}$. This gives $Z/Z_{\odot} = 0.02^{+0.02}_{-0.01}$, or $Z=(1-4)$\% $Z_{\odot}$ (Table 3). 
 
With an estimate of the equivalent hydrogen column density of the absorber in hand, we can estimate the thickness of the absorber along the line of sight. This is given, assuming homogeneity by the ratio between the column and the volume baryon densities of the absorber: $D=$N$_b/$n$_b$. 
The value of n$_b$ is not constrained by the XMM-{\em Newton} data of PKS~0558-504. However, hydrodynamical simulations that include feedback from galaxy super-winds predict that $n_b$ in the WHIM correlates with temperature (e.g. Figure 3 in Cen \& Ostriker, 2006), and that for a given  temperature, lower metalicities are normally found in higher density environments (e.g. Figure 14 in Cen \& Ostriker, 2006)
\footnote{This is because super-winds are very efficient in metal-polluting the IGM in the proximity of the host galaxy, but can only spread  a finite amount of metals in a finite volume. Therefore IGM regions with lower primordial baryon densities will increase their gas metal content more than regions with initially higher baryon densities.}
. From Fig. 14 of Cen \& Ostriker (2006), at $T\sim 5 \times 10^6$ K, metalicites of the order of log$(Z/Z_{\odot}) \simeq -1.5$ are reached at overdensities, compared to the average density of the Universe ($<n_b>= 2 \times 10^{-7}  (1+z)^3 (\Omega_bh^2/0.02)$) of $\delta = n_b/<n_b> \simeq 300$
\footnote{This is consistent with our a-priory assumption of freezing the value of the baryon density to $n_b=10^{-5}$ cm$^{-3}$ in our fit to the X-ray data, since at $n_b \ge 10^{-5}$ cm$^{-3}$ the contribution of photoionization is negligible, and the best-fitting temperature and column densities are not affected by the exact value of this paremeter.}
. At $z=0.118$ and assuming $h=0.7$, and $\Omega_b=0.046$, this gives $n_b \simeq 10^{-4}$ cm$^{-3}$. 
This implies $D \simeq 5_{-1}^{+2} \times (n_b/<n_b>)^{-1}$ Mpc. 

Table 3 summarize the physical and geometrical parameters of the WHIM filament at $z=0.118 \pm 0.001$, along the line of sight to PKS~0558-504. 

\subsection{Number Density and Cosmological Mass Density of WHIM}
A tentative OVIII detection along a single line of sight cannot be used to measure the number density of OVIII filaments and the cosmological mass density in the WHIM. However, we think that, in this particular case, this exercise is instructive, because it clearly shows the serendipitous nature of the detection, and its bias toward dense (i.e. high equivalent width) and rare absorbers. 

We estimate the number density of OVIII WHIM filaments, $dN_{OVIII}/dz$ and $\Omega_b^{WHIM}$, by taking into account the large Poissonian errors associated with small number statistics (Gehrels, 1986)
\footnote{We do not try to include systematics here, because of the aready big (and probably domaniting) statistical uncertainties associated with a single detection, which allow for the estimate to be consistent with all possible theoretical predictions.}
. We get: $dN_{OVIII}/dz = 7.3_{-6.6}^{+21.1}$ and $\Omega_b^{WHIM} = 0.18_{-0.16}^{+0.54}$ (90\% significance). These numbers should be compared to the expected number density of WHIM filaments along a random line of sight, and to the total number of baryons as inferred by both microwave background anysotropies ($\Omega_b = 0.046 \pm 0.002$, Bennett et al., 2003; Spergel et al., 2003) and by the standard ``big-bang nucleosynthesis'' when combined with light-element ratios ($\Omega_b = 0.044 \pm 0.004$, Kirkman et al., 2003). From Cen \& Fang (2006; their Figure 4), we derive $dN_{OVIII}/dz(EW>9$ m\AA$) \simeq 0.8$. This is consistent with the 90\% significance lower limit of our estimate. Similarly, our estimate of $\Omega_b^{WHIM}/\Omega_b = 3.9^{+11.7}_{-3.5}$ greatly exceeds the expected value of about 40-50 \% at $z\sim 0$ (e.g. CO06), but is consistent within the large 90\% uncertainities.
This is not surprising, since the WHIM detection we report is a serendipitous discovery in the spectrum of a target whose observations 
were not planned with this aim, and as such is biased toward dense (i.e. high equivalent width) and rare absorbers. 

\subsection{Caveats and their Solutions}
Although the association between UV and X-ray absorbers reported here, makes this detection of a probable WHIM filament the first for which a metallicity measure is possible, we caution that, due to instrumental limitations and uncertainties, a definitive identification of this absorber with an intervening WHIM filament, is not possible with the current data. 

First, the absorption that we identify here in X-ray as an OVIII Ly$\alpha$ at $z=0.117 \pm 0.001$, could also, in principle be identified as OVII K$\beta$ near the systemic redshift of PKS~0558-504. We reject this interpretation based on the non-detection of associated OVII K$\alpha$ absorption with the predicted strength ($\gs 6$  times stronger than the detected line) - together with several other strong lines that should be detectable in the X-ray spectrum (CVI, NVI, NVII, NeIX, etc.) - and the lack of intrinsic OVI absorption in the RGS and FUSE spectra of PKS~0558-504, respectively. Moreover, the absorption seen both in FUSE and IUE and here identified as HI absorption cannot plausibly be  identified with any transition at or near the systemic redshift of the target. Unfortunately, however, the OVII K$\alpha$ transition which would be associated with OVII K$\beta$ at $z\simeq z_{sou}$, falls close to a strong RGS instrumental feature, which makes the source-frame absorption hypothesis difficult to reject with high confidence. The identification of the reported X-ray absorber with intrinsic nuclear OVII K$\beta$ is thus unlikely, but not impossible based on the current X-ray data only. 
The only way of confirming the intervening origin of the X-ray absorber would be to clearly detect the other two lines predicted by the best-fitting WHIM model (one of which already hinted in the GO RGS1 and CAL RGS2 spectra of PKS~0558-504): i.e. FeXVII L($\lambda=15.015$ \AA) and NeIX K$\alpha$, which fall at $\lambda=15.02$ and $\lambda=16.77$ \AA. This is a task that only the {\em Chandra} LETG grating could accomplish, thanks to its $40$ cm$^2$ effective area at this energies (vs $\sim 50$ cm$^2$ for the RGS2 and $\sim 15$ cm$^2$ for the {\em Chandra} MEG), and factor of $\sim 2.5$ better spectral resolution, compared to the RGS. 

Second, while our identification of the absorber as a system at $z=0.118 \pm 0.001$ is the most probable with the currently available data, its origin might not be intervening. The absorber could also be intrinsic to the nucelar Seyfert environment, outflowing from the nucleus of PKS~0558-504. We discuss this possibility in \S 5.1, and conclude that the physical parameters inferred for this absorber, make this interpretation unlikely. A way to definitely exclude this possibility, is to look for variability of the X-ray and far-UV features reported here. Observations with the HST-COS of the far-UV spectrum of PKS~0558-504, would not only definitely confirm or rule out the existence of Ly$\alpha$-$\beta$ absorption at $z=0.1183 \pm 0.0001$ (the redshift in FUSE), but would also test for opacity (i.e. ionization state and/or column density) variability of this absorber, which, if detected, would make the WHIM identification impossible. Analogously, new high S/N X-ray ({\em Chandra}) high-resolution spectra of PKS~0558-504 would allow one to check for variability of the X-ray absorber. 

A third observation that can help sheding light on the true identification of this absorber, by either strongly supporting or weakening the support for its intervening nature, is a measure of the galaxy density in the region surrounding the putative WHIM filament at $z=0.118$. WHIM filaments are supposed to correlate strongly with concentrations of galaxies and with Large Scale Structure (e.g. Stocke et al., 2006b). Unfortunately, the line of sight to PKS~0558-504 lies in the Southern sky, at a location which is currently poorly investigated: none of the major galaxy surveys or deep surveys (e.g. 2dF
\footnote{http://www.mso.anu.edu.au/2dFGRS/}
, 6dF
\footnote{http://www.aao.gov.au/local/www/6df/}
, SDSS
\footnote{http://www.sdss.org/}
), covers this region of the sky. 
Wide-field, multi-band observations, centered on the line of sight to PKS~0558-504, and sensitive down to a fraction of L$^*$ at $z=0.118$, would allow photometric redshifts of the galaxies in the field to be determined with sufficient accuracy to exclude foreground and background objects. Spectroscopic follow-ups would then allow one to precisely estimate the galaxy density at $z=0.118$ in a surrounding area around the line of sight to PKS~0558-504, to investigate on the association of WHIM filaments and LLSs. 

\section{Conclusions}
We reported on the first combined and possibly associated Far-UV and X-ray tentative detection of a WHIM filament at a mean common redshift of $z=0.118 \pm 0.001$, along the line of sight to the Seyfert PKS~0558-504. 
Our main findings are: 

\begin{itemize}

\item{} OVIII Ly$\alpha$ absorption is tentatively identified at $z=0.117 \pm 0.001$ in two independent XMM-{\em Newton} RGS1 spectra of the Seyfert galaxy PKS~0558-504: a high S/N 480 ks net GO spectrum taken in 2008, and the sum of a number of much lower S/N spectra taken between 2000-2001 as part of an XMM-{\em Newton} calibration campaign, with a total net expsoure of 309 ks. The combined, single-line significance of these two detections is $2.8\sigma$. 

\item{} When fitted with a model including proper parameterizations of the continua together with our hybrid (collisional ionization plus photoionization) WHIM absorption model, the two ful-band RGS1 and RGS2 spectra are consistent with the presence of a WHIM filament at $z=0.117 \pm 0.001$ with logT$=6.56_{-0.17}^{+0.19}$ and N$_H = (3.2^{+3.1}_{-1,6}) \times 10^{19} (Z/Z_{\odot})^{-1}$ cm$^{-2}$. Based on X-ray data only, the parameters of this model are poorly constrained and both represent only lower limits at a $3\sigma$ statistical confidence level. 

\item{} Archival FUSE data of PKS~0558-504 show the presence of a broad absorption complex at $\lambda= 1147.1 \pm 0.1$, with single-line statistical significance of $4.1\sigma$ ($3.7\sigma$ when systematics are included). This line can be identified as broad HI Ly$\beta$ at $z_{BLB}=(0.1183 \pm 0.0001)$, consistent with the redshift of the putative X-ray OVIII Ly$\alpha$, within the large X-ray 
uncertainties. The width of this BLB absorber is marginally consistent, at a 2.5$\sigma$ level, with the thermal width of HI ions in gas with logT$=6.56_{-0.17}^{+0.19}$ (the best-fit temperature derived from the X-ray data). 

\item{} Archival IUE-SWP, low-resolution, spectra of PKS~0558-504 taken in 1987 and 1989, also hint to the presence of an unresolved intervening HI Ly$\alpha$ at $z_{BLA}=0.119 \pm 0.001$, with a single-line statistical significance of only $1.7\sigma$. The redshift of this absorber is consistent, within their relative 1$\sigma$ uncertainties, with the redshift $z_{BLB}$ of the HI absorber detected with FUSE. The BLA (IUE) to BLB (FUSE) equivalent width ratio of these two lines, is consistent with the expected unsaturated value. 

\item{} We tentatively associate the OVIII Ly$\alpha$ X-ray absorber at $z=0.117 \pm 0.001$, with the far-UV BLB and BLA absorbers at $z_{BLB}=(0.1183 \pm 0.0001)$ and $z_{BLA}=0.119 \pm 0.001$, respectively, and identify them with an intervening WHIM filament at the mean common redshift of $z=0.118 \pm 0.001$. The combined significance of this detection is $5.2\sigma$ ($4.6\sigma$ if 
systematics in the FUSE continuum modeling are taken into account and the IUE HI BLA line is not considered). 

\item{} The above identification allows us to estimate for the first time the equivalent hydrogen column density of the absorber and so its metallicity under the assumption that the bulk of the HI BLB and the OVIII absorbers are physically associated. These turn out to be: N$_H = (1.5_{-0.4}^{+0.7}) \times 10^{21}$ cm$^{-2}$ and $Z=(1-4)$\%$Z_{\odot}$. If the HI BLB and the OVIII absorbers are not directly 
associated or are structured, the above estimate represent upper and lower limits, respectively. 

\item{} The non-detection of associated OVI absorption in the FUSE spectrum of PKS~0558-504, allows us to put a stringent lower limit on the temperature of the absorber: by combining X-ray and FUSE data we obtain: logT$=(6.56_{-0.04}^{+0.19})$ K. 

\item{} From the theoretical correlation between the temperature of WHIM filaments and their overdensity (relative to the average density in the Universe) and the expected 3D metallicity-temperature-overdensity relationship, we derive an overdensity of $\delta\sim 300$ for our system, and therefore a thickness along the line of sight of $5_{-1}^{+2}$ Mpc. 

\item{} Finally from this single detection we extrapolate the number density of OVIII Ly$\alpha$ WHIM absorbers, and the cosmological mass density of WHIM. Both are consistent, within their large $1\sigma$ uncertainties due to the low-number statistics, with the expected values, but their central-values are significantly higher due to the natural bias toward dense (i.e. high equivalent width) and rare absorbers associated with  serendipitous WHIM detections. 

\end{itemize}

\begin{acknowledgments}
FN, LZ and MLC acknowledge support from the LTSA grant NNG04GD49G. FN acknoleges support from the FP7-REGPOT-2007-1 EU grant No. 206469.
This research has made use of the NASA/IPAC Extragalactic Database (NED) which is operated by the Jet Propulsion Laboratory, California Institute of Technology, under contract with the National Aeronautics and Space Administration.
\end{acknowledgments}


%
\begin{table}
\footnotesize
\begin{center}
\caption{\bf Log of the XMM-{\em Newton}, FUSE and IUE Observations}
\vspace{0.4truecm}
\begin{tabular}{|c|ccc|}
\hline
Dataset & Date of Obs. & Obs. ID & Exposure$^a$ \\
\hline
XMM-RGS CAL & 2000/02/07 & 0116700301$^b$ & 19.0 \\   
& 2000/02/10 & 0117500201 & 40.8 \\ 
& 2000/02/12 & 0117710701 & 49.9\\ 
& 2000/02/13 & 0117710601 & 56.9\\  
& 2000/02/14 & 0117710501 & 6.8 \\     
& 2000/03/07 & 0120300501 & 6.5 \\
& 2000/03/07 & 0120300601 & 7.0 \\  
& 2000/03/07 & 0120300801 & 39.7  \\ 
& 2000/05/24 & 0125110101 & 30.0 \\  
& 2000/10/10 & 0129360201 & 26.4 \\  
& 2001/06/26 & 0137550201 & 14.8 \\  
& 2001/10/19 & 0137550601 &  14.6 \\
\hline  
XMM-RGS GO & 2008/09/07 & 0555170201 & 121.4 \\
& 2008/09/09 & 0555170301 & 127.8 \\
& 2008/09/11 & 0555170401 & 126.2 \\
& 2008/09/13 & 0555170501 & 126.6\\
& 2008/09/13 & 0555170601 & 115.8 \\
\hline
FUSE-LWRS & 1999/12/10 & P1011504000 & 46.2 \\
& 2002/11/07 & C1490601000 & 48.3 \\
\hline
IUE-SWP LR & 1987/09/22 & SWP31899 & 13.8 \\
& 1989/11/14 & SWP37589 & 13.8 \\
& 1989/11/15 & SWP37604 & 16.2 \\
\hline
\end{tabular}
\end{center}
$^a$ Net, in ksec. 
$^b$ RGS1 only. 
\end{table}
\normalsize

\begin{table}
\footnotesize
\begin{center}
\caption{\bf Absorption Lines (and upper limits) in the XMM-{\em Newton}  RGS, FUSE and IUE spectra of PKS~0558-504}
\vspace{0.4truecm}
\begin{tabular}{|cccc|cc|}
\hline
Wavelength & Width & EW & Significance & Id & Redshift\\
in \AA\ & FWHM in km s$^{-1}$ & in m\AA\ & in $\sigma$ & & in km s$^{-1}$ if z$<10^{-3}$ \\ 
\hline
\multicolumn{4}{|c|} {XMM-{\em Newton} RGS} & & \\
$23.52 \pm 0.03$ & $<900$ & $23.7_{-3.7}^{+5.0}$ & 6.4 & OI$_{K\alpha}$ & $\pm 380$ km s$^{-1}$ \\
$21.61 \pm 0.03$ & $<900$ & $12.9 \pm 5.2$ & 2.5 & OVII$_{K\alpha}$ & $138 \pm 380$ km s$^{-1}$ \\
$21.17 \pm 0.03$ & $<1200$ & $9.1 \pm 3.3$ & 2.8 & OVIII$_{K\alpha}$ & $0.116 \pm 0.002$ \\
15.00-15.03$^a$ & $<900$ (fixed) & $< 3.0$ & 1$^b$ & NeIX$_{K\alpha}$ & $0.117 \pm 0.001$ \\
16.76-16.79$^a$ & $<900$ (fixed) & $< 5.0$ & 1$^b$ & FeXVII$_{L}$ & $0.117 \pm 0.001$ \\
\hline
\multicolumn{4}{|c|} {FUSE-LWRS} & & \\
$1147.1 \pm 0.1$ & $160^{+50}_{-30}$ & $66 \pm 16$ & 4.1 & HI$_{Ly\beta}$ & $0.1183 \pm 0.0001$ \\
1152.7-1154.7$^a$ & 50 (fixed) & $< 12$ & 1$^b$ & OVI($\lambda=1031.9$) & $0.118 \pm 0.001$ \\
\hline
\multicolumn{4}{|c|} {IUE-SWP LR} & & \\
$1360 \pm 1$ & $<1300$ & $1000 \pm 600$ & 1.7 & HI$_{Ly\alpha}$ & $0.119 \pm 0.001$\\
\hline
\end{tabular}
\end{center}
$^a$ Wavelength interval over which the 1$\sigma$ EW upper limit is computed, corresponding to $z=0.117 \pm 0.001$ for the X-ray lines, and $z=0.118 \pm 0.001$ for the UV line. 

$^b$ 1$\sigma$ upper limit. 
\end{table}
\normalsize

%
\begin{table}
\footnotesize
\begin{center}
\caption{\bf Physical and Geometrical Parameters of the WHIM Filament at $z=0.118 \pm 0.001$}
\vspace{0.4truecm}
\begin{tabular}{|c|cccc|}
\hline
Redshift & log(T) &  N$_H$ & Metallicity & Thickness $^a$ \\
& (T in K) & (in 10$^{21}$ cm$^{-2}$) & & in Mpc \\
\hline
$0.118 \pm 0.001$ & $6.56^{+0.19}_{0.04}$ & $1.5_{-0.4}^{+0.7}$ & (1-5) \% $Z_{\odot}$ & $5_{-1}^{+2}$ \\
\hline
\end{tabular}
\end{center}
$^a$ For a baryon volume density of n$_b = 10^{-4}$ cm$^{-3}$. 

\end{table}
\normalsize

\begin{figure}
\plotone{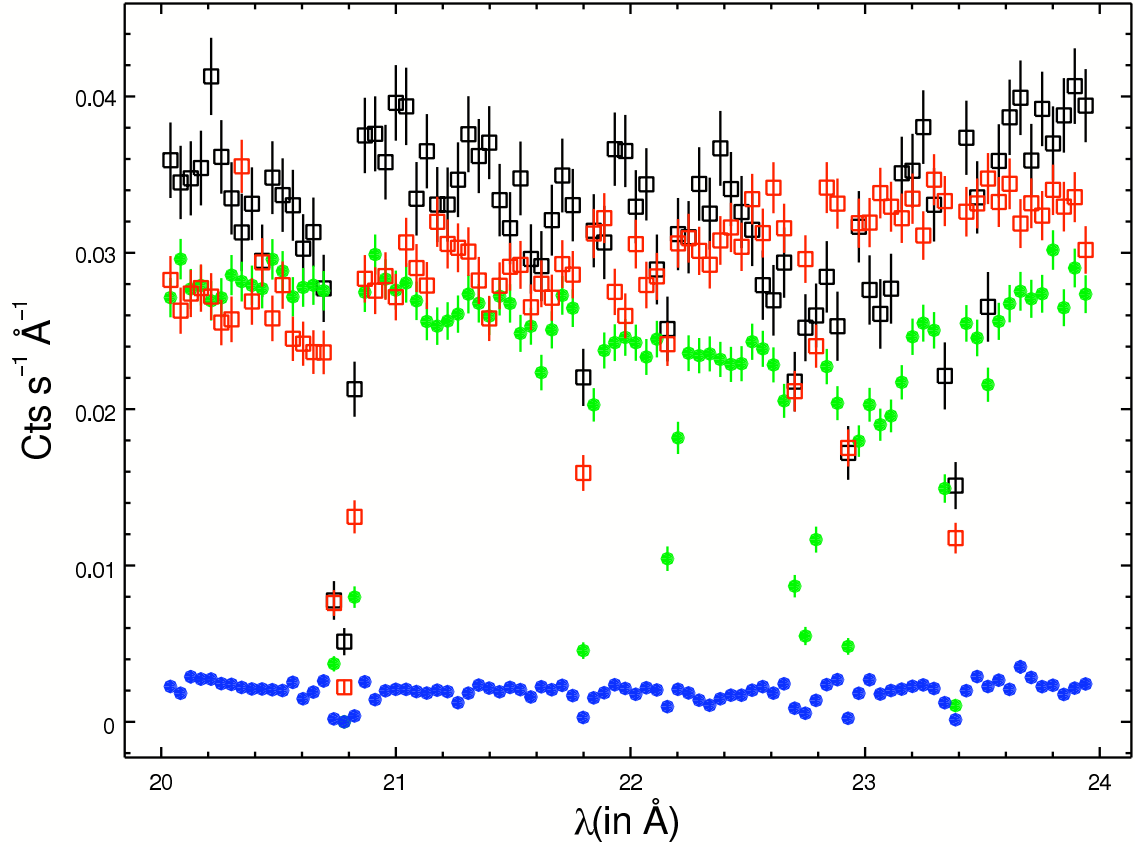}
\caption{20-24 \AA\ portion of the RGS1 CAL (empty squares) and GO (filled circles) source (black and green , respectively) and background 
(red and blue, respectively) spectra of PKS~0558-504. The average background level during the CAL observations is about 10 times 
higher than that of the GO observations.
\label{fig1}}
\end{figure}

\begin{figure}
\plotone{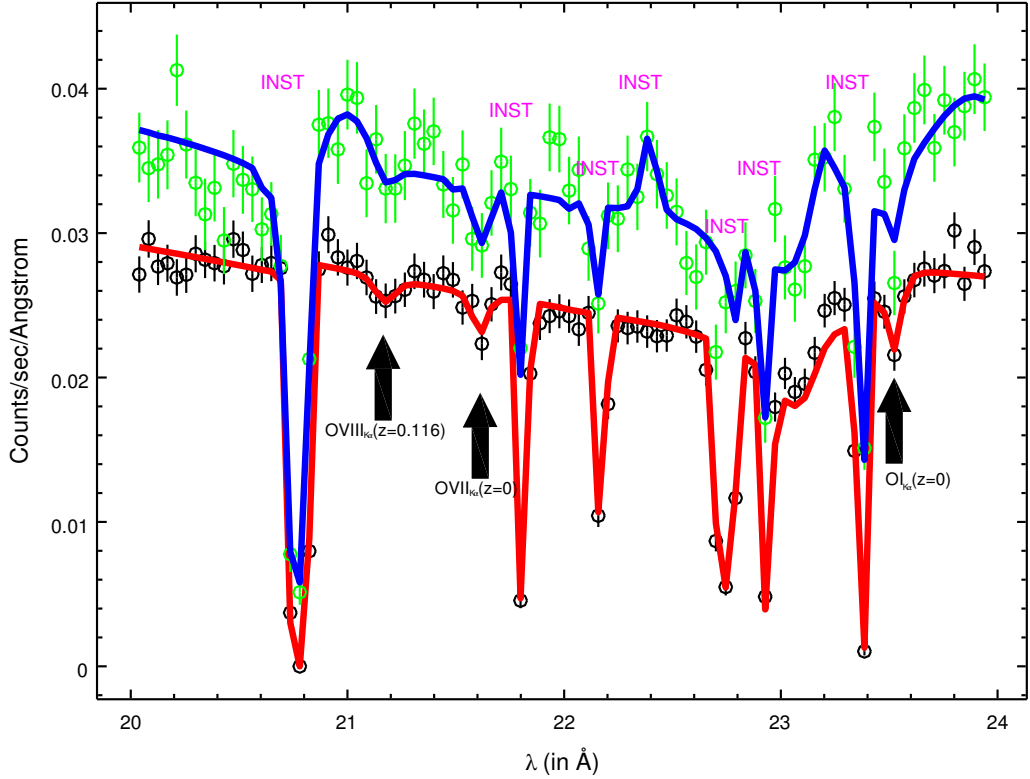}
\caption{20-24 \AA\ portion of the CAL (green points and errorbars) and GO (black points and errorbars) RGS1 spectra of PKS~0558-504, along 
with their best fitting continuum plus absorption models. The deep line-like features present in both spectra are instrumental features, due to bad-pixels/columns of the read out CCD. The remaining three real features present in both spectra are marked and here identified as 
OI K$\alpha$ and OVII K$\alpha$ at $z\simeq 0$, and, tentatively, as OVIII Ly$\alpha$ at $z=0.116 \pm 0.002$. 
\label{fig2}}
\end{figure}

\begin{figure}
\plotone{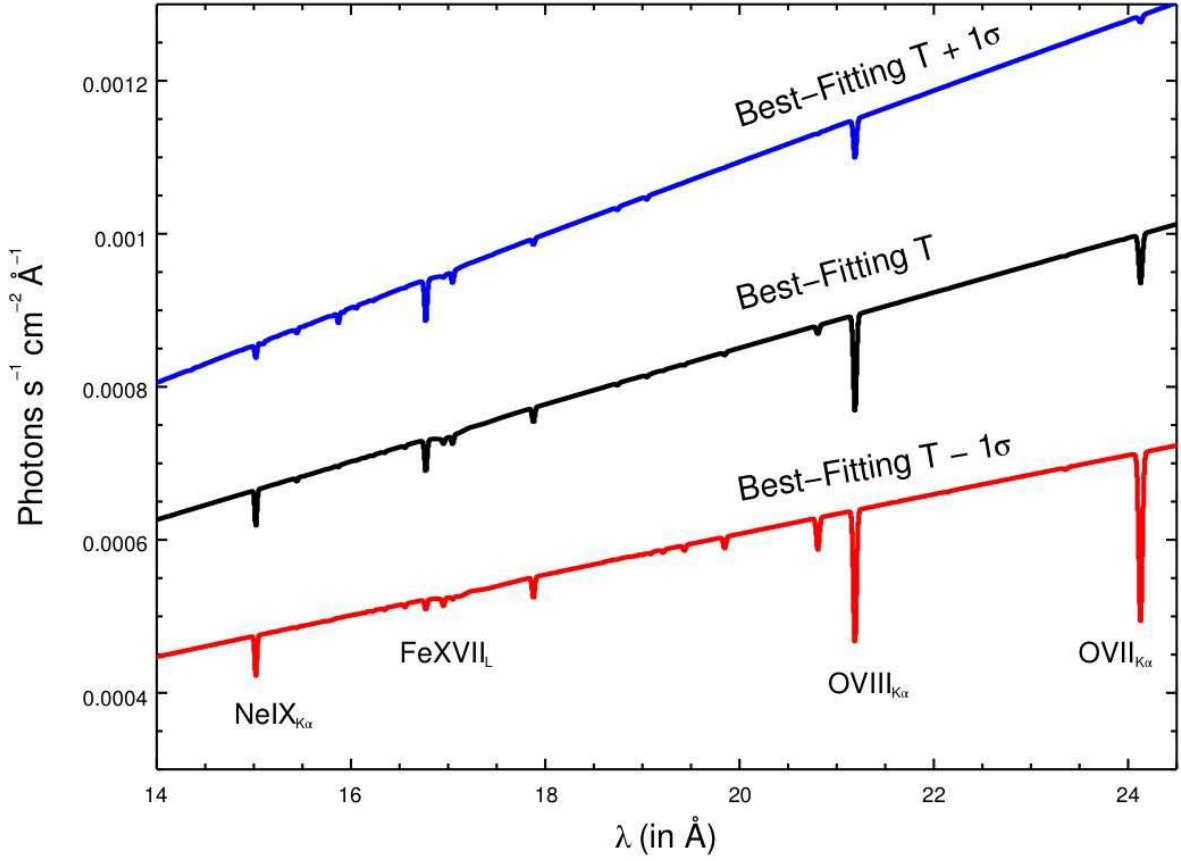}
\caption{Best fitting WHIM model (black curve) together with its negative (red curve) and positive (blue curve) 1$\sigma$ temperature 
indeterminations. 
\label{fig4}}
\end{figure}

\begin{figure}
\plotone{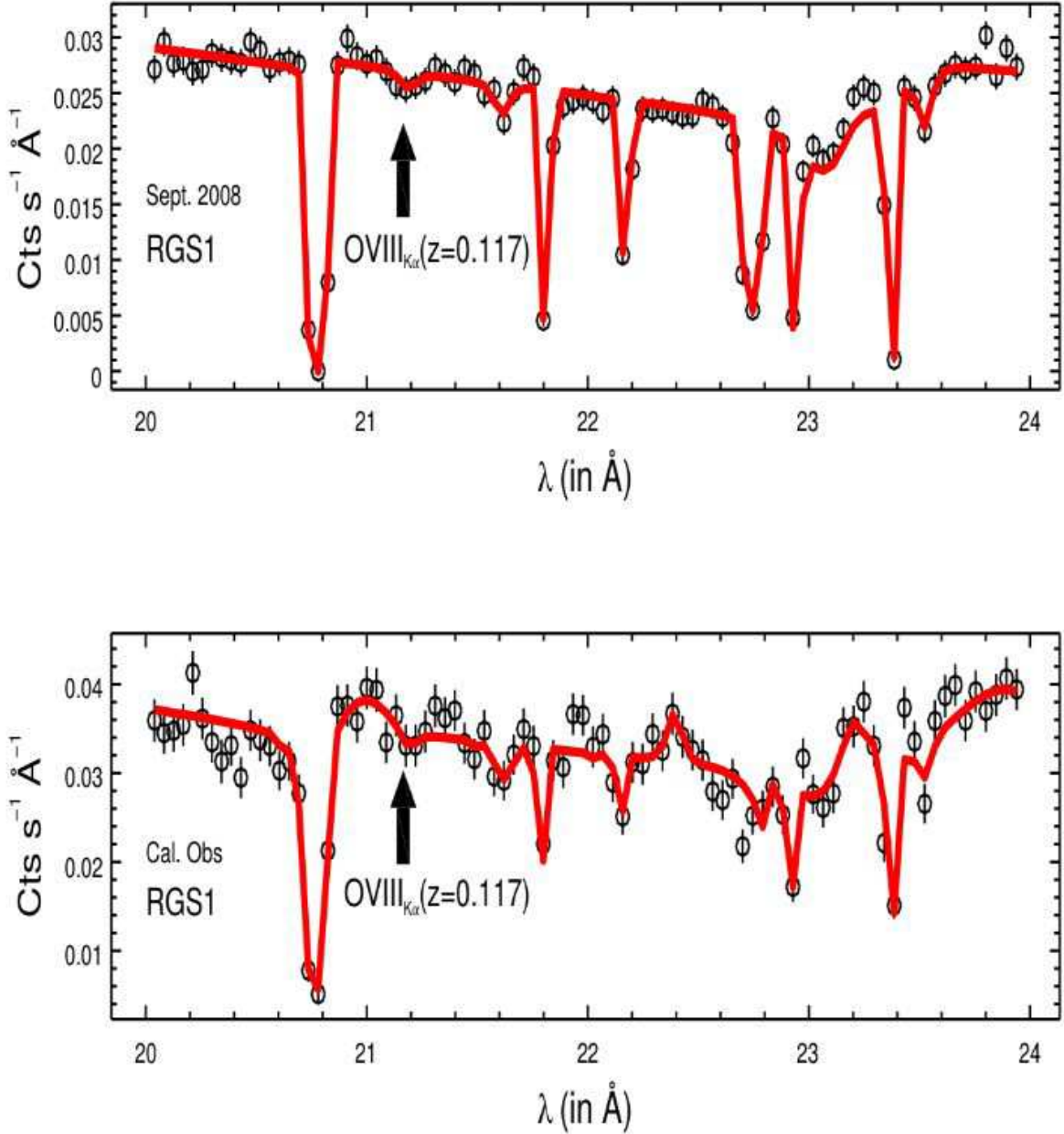}
\caption{21-24 \AA\ portion of the RGS1 GO (top panel) and CAL (bottom panel) spectra of PKS~0558-504. 
The superimposed red curve is the best fitting WHIM model convolved with the instrumental responses. 
\label{fig3}}
\end{figure}

\begin{figure}
\plotone{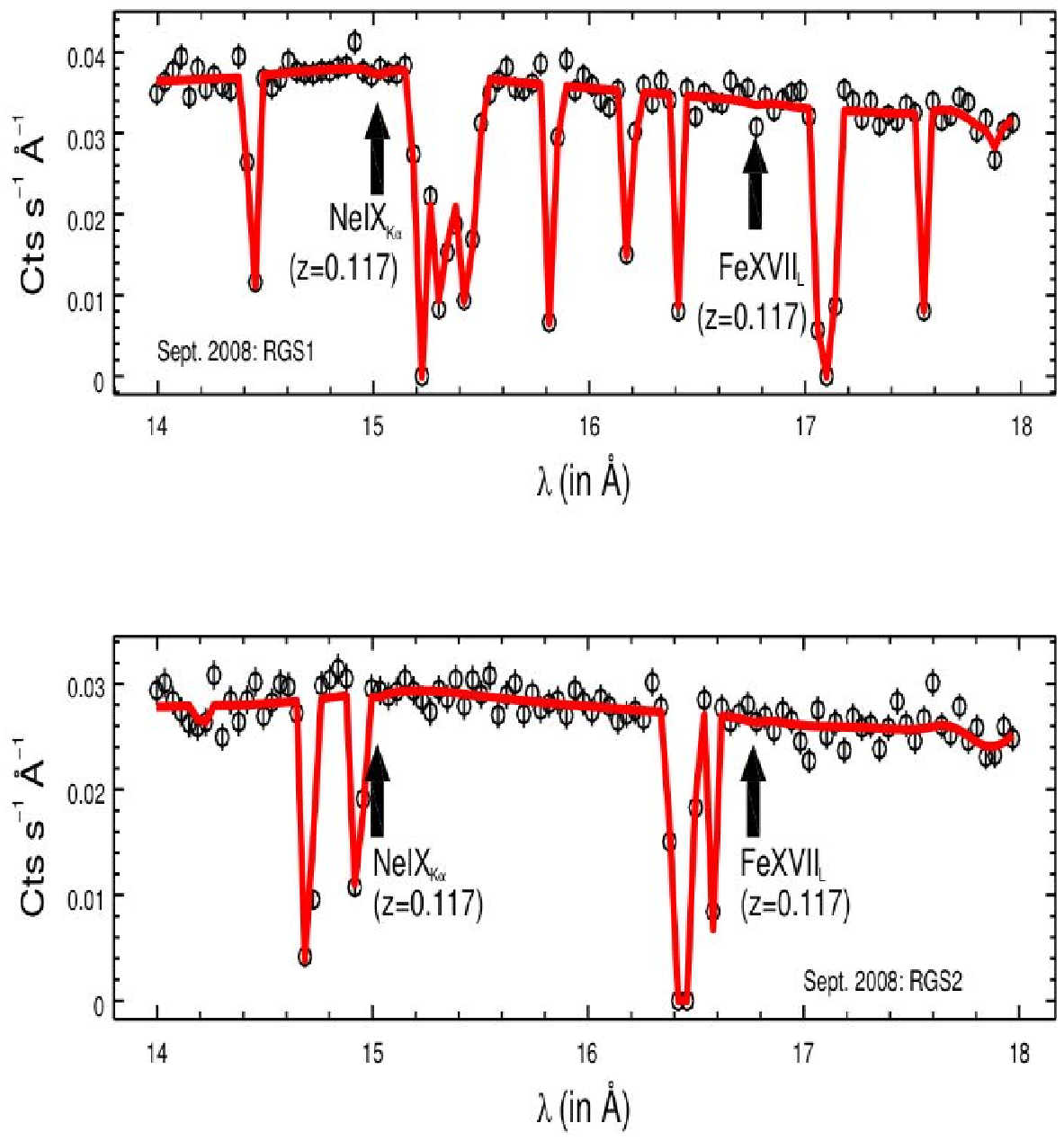}
\caption{14-18 \AA\ portion of the RGS1 (top panel) and RGS2 (bottom panel) GO spectra of PKS~0558-504. 
The superimposed red curve is the best fitting WHIM model convolved with the instrumental responses.
\label{fig3}}
\end{figure}

\begin{figure}
\plotone{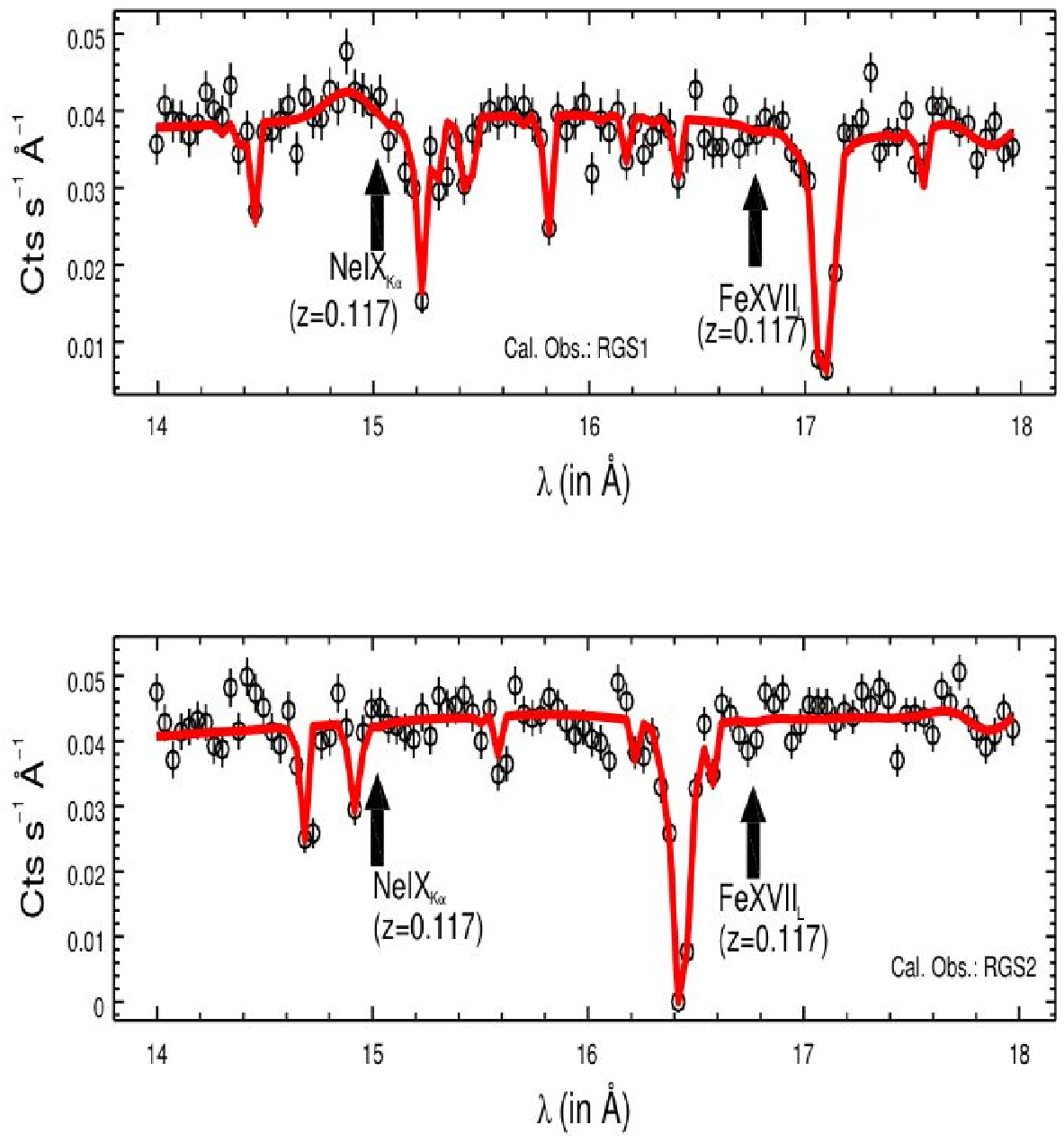}
\caption{14-18 \AA\ portion of the RGS1 (top panel) and RGS2 (bottom panel) CAL spectra of PKS~0558-504. 
The superimposed red curve is the best fitting WHIM model convolved with the instrumental responses. 
\label{fig3}}
\end{figure}

\begin{figure}
\plotone{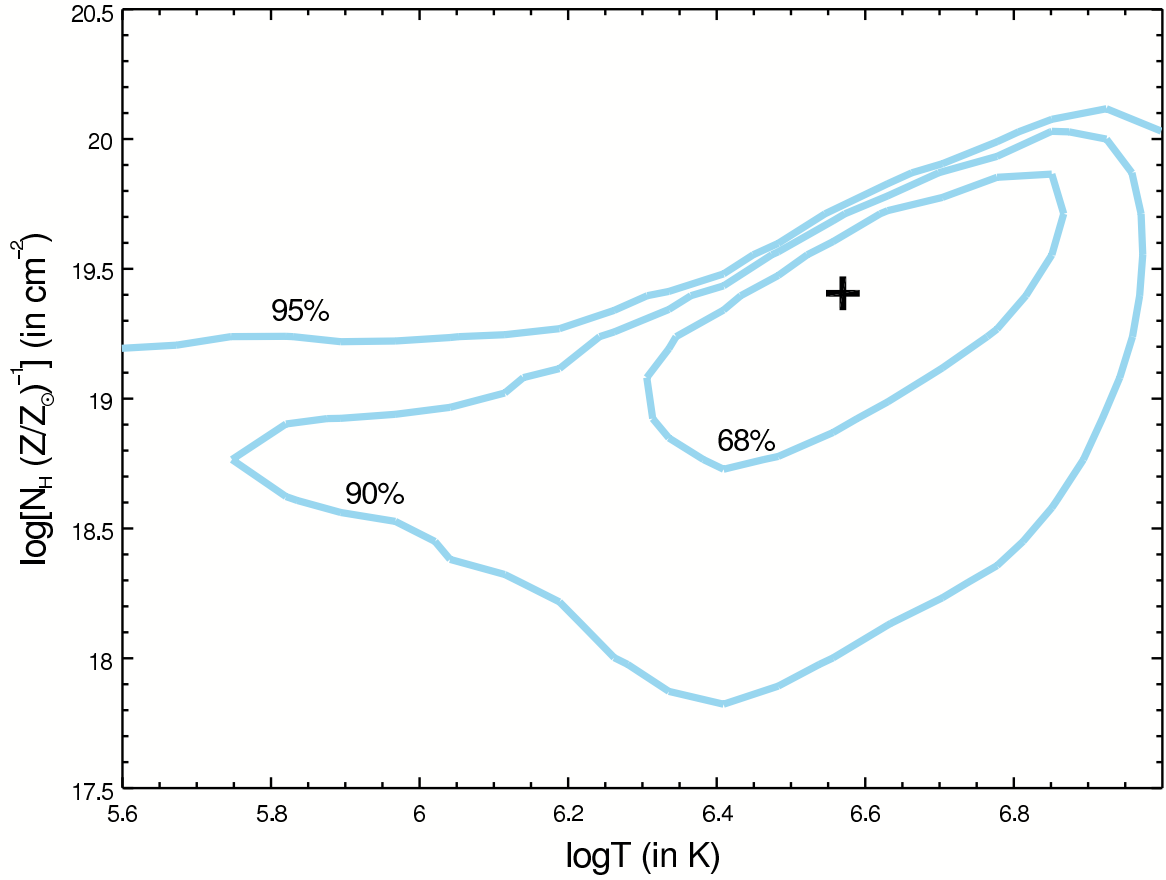}
\caption{68\%, 90\% and 95\% log$T$ - logN$_H$ contours for the putative X-ray WHIM filament at $z=0.117 \pm 0.001$
\label{fig5}}
\end{figure}

\begin{figure}
\plotone{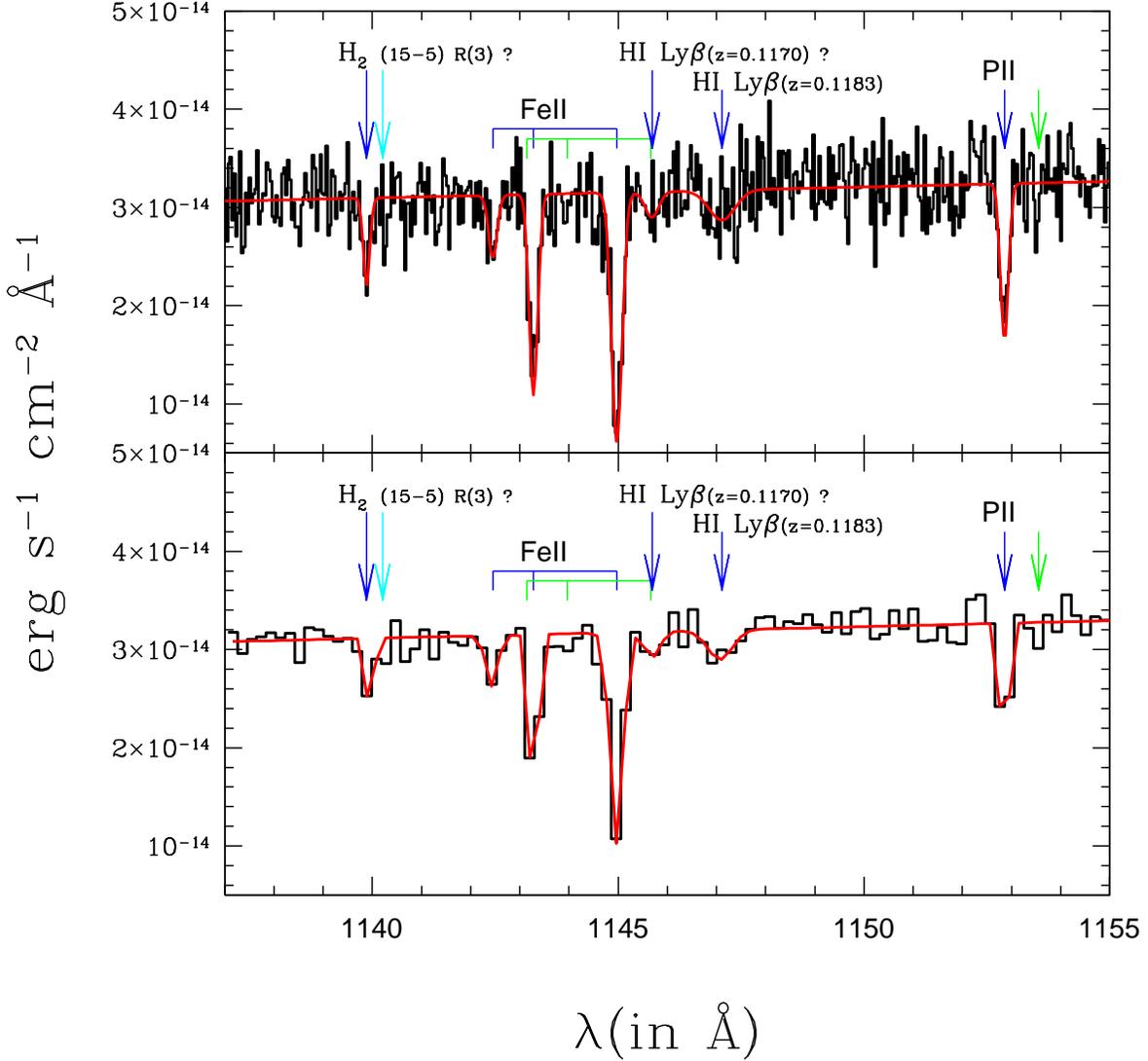}
\caption{1137-1155 \AA\ portion of the FUSE spectrum of PKS~0558-504, in two different binning schemes: $\sim 15$ km s${^-1}$, the FUSE 
resolution (top panel), and $\sim 90$ km s$^{-1}$ (bottom panel). Absorption lines identifications are labeled in both panels. 
Low- and high-velocity components of the low-ionization Fe and P and molecular H$_2$absorbers are marked in blue and green and blue and cyan, respectively, but only the much stronger low-velocity lines are fitted (with the exception of the high-velocity FeII($\lambda 1144.94$) line at $\lambda = 1145.7$ \AA, which we fit because possibly partly contamined by intervening HI absorption). 
The broad absorption complex that we tentatively identify as intervening HI Ly$\beta$ is clearly better seen in the heavily binned spectrum (bottom panel). 
\label{fig6}}
\end{figure}

\begin{figure}
\plotone{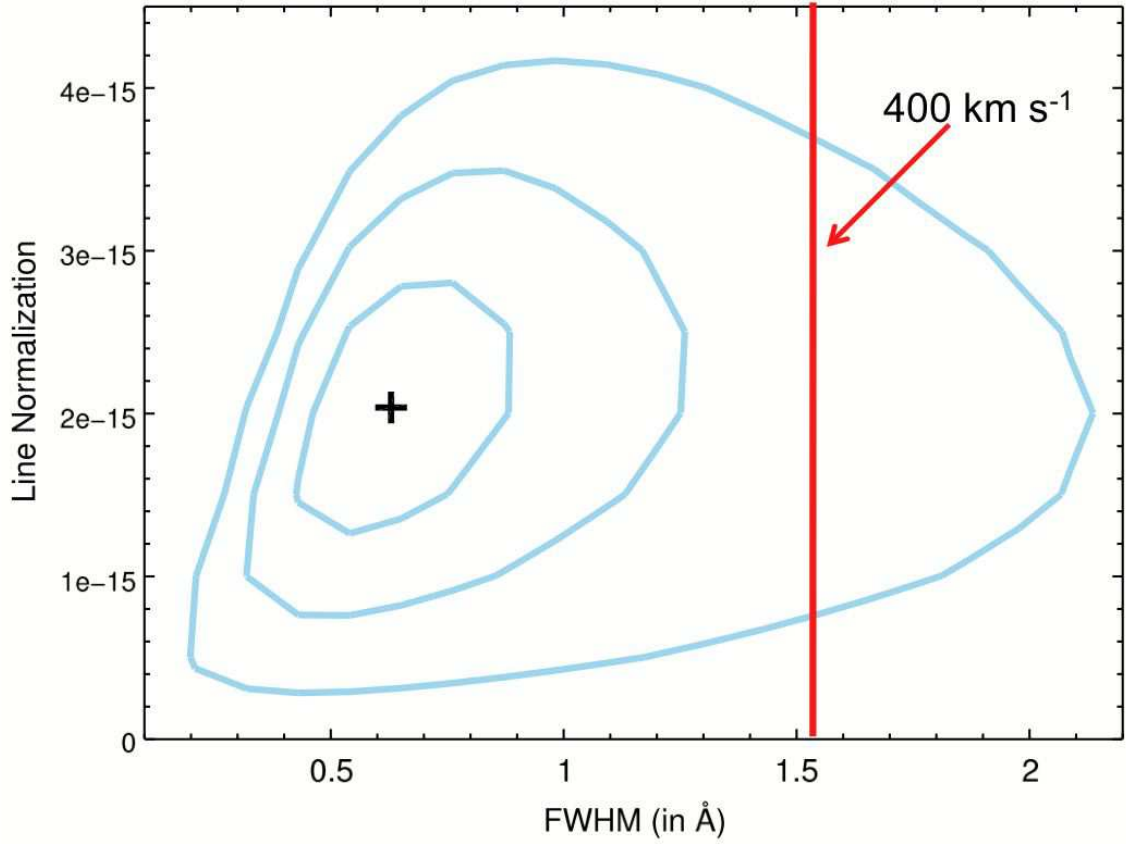}
\caption{Gaussian normalization and FWHM 1, 2 and 3$\sigma$ contour plot, for the $z_2=(0.1183 \pm 0.0001)$ HI Ly$\beta$ 
absorber. FWHM=1.5 \AA\ corresponds to $v=390$ km s$^{-1}$ at $\lambda=1147.1$ \AA, and so to the expected doppler 
parameter of $b\simeq 165$ km s$^{-1}$, for hydrogen in gas at logT$\sim 6.5$. 
\label{fig6}}
\end{figure}

\begin{figure}
\plotone{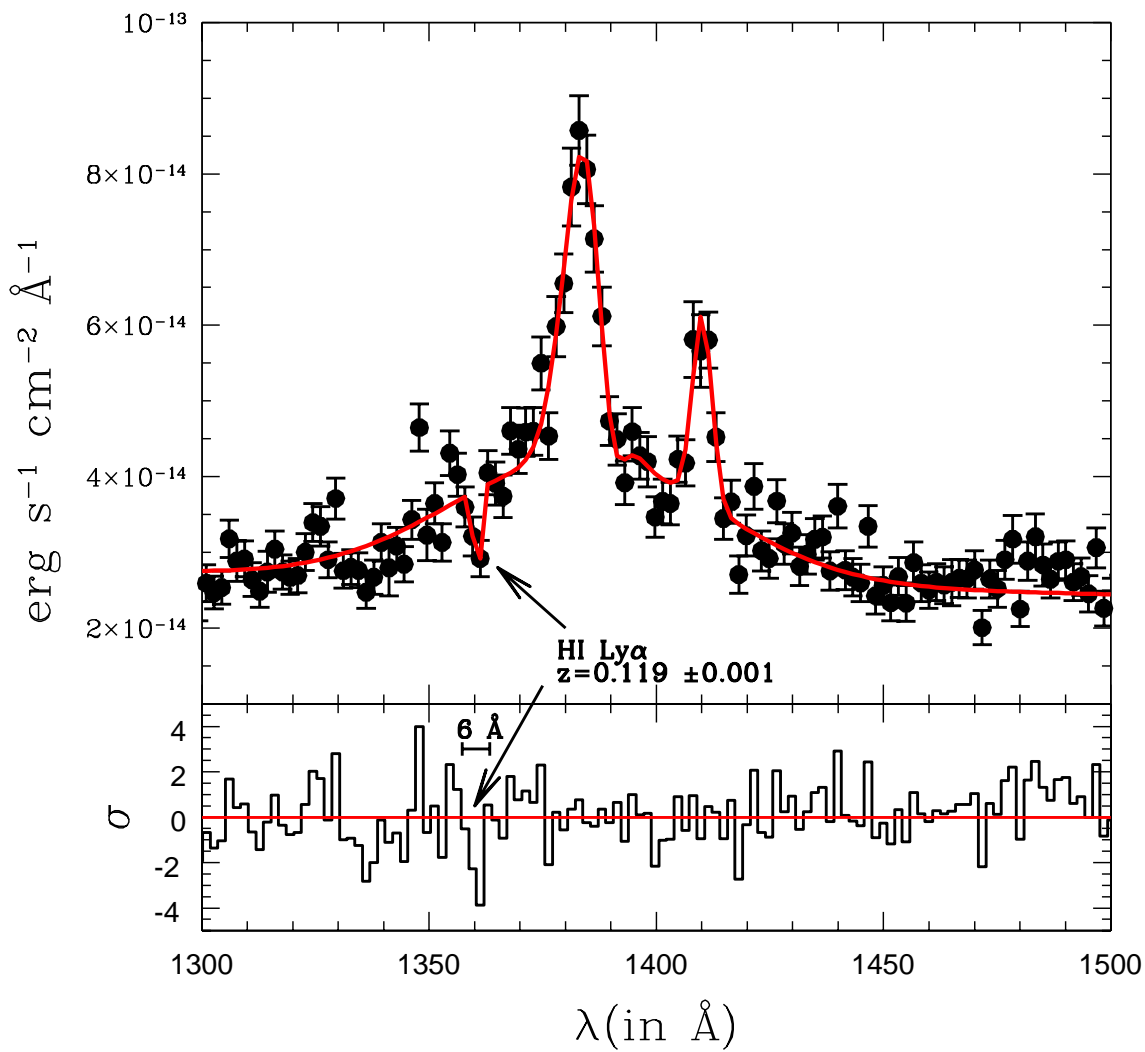}
\caption{1300-1500 \AA\ portion of the 1989, November 15 IUE-SWP LR spectrum of PKS~0558-504 (obsid 37604, Table 1) with our best-fitting superimposed (top panel), and the residuals after subtracting the Gaussian at $\lambda=1360 \pm 1$ \AA\ (bottom panel). Our tentative identification of a HI Ly$\alpha$ absorber at $z=0.119 \pm 0.001$, is labeled. 
\label{fig6}}
\end{figure}

\begin{figure}
\plotone{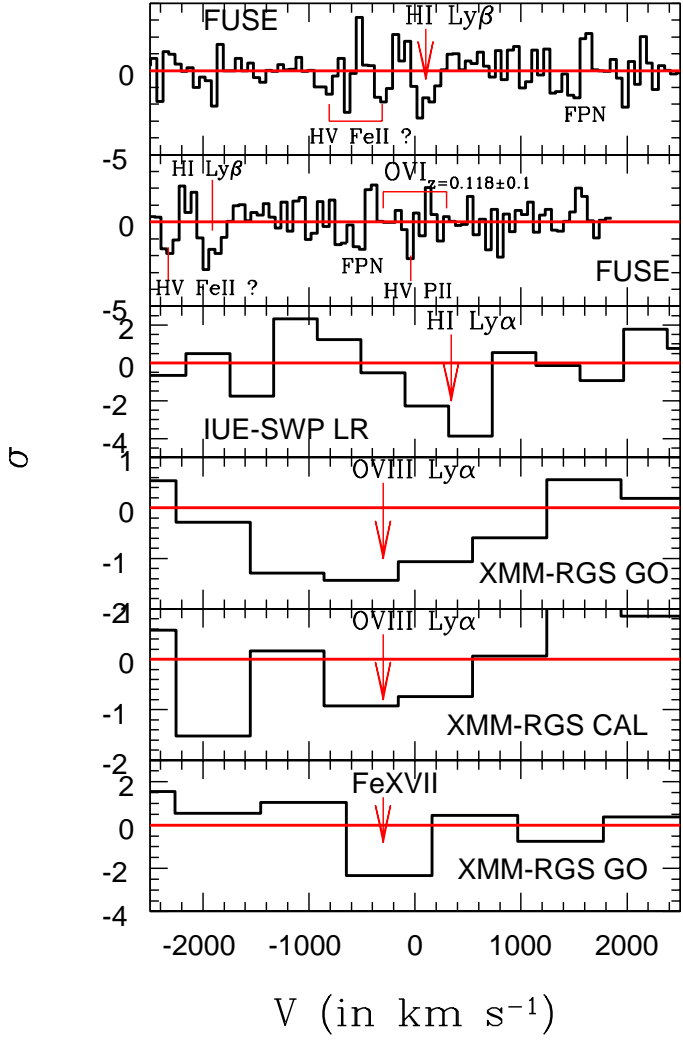}
\caption{Residuals (in $\sigma$), in velocity space, to the best fitting continuum (plus ISM low-velocity lines, in the case of FUSE) models to the FUSE (first and second panel from the top), IUE-SW LR (1989; third panel from the top) and XMM-{\em Newton} (GO: 4rd and 6th panels from the top; CAL: 5th panel from the top) spectra of PKS~0558-504. The residuals are centered around the absorption lines that we identify as HI Ly$\beta$, HI Ly$\alpha$, OVIII Ly$\alpha$ and FeXVII L ($\lambda=15.015$ \AA, rest frame), at a mean common redshift of $z=0.118 \pm 0.001$. The red arrows in the figures indicate the best-fitting centroid of each line in their respective spectra. The red segment labeled ``OVI$_{z=0.118\pm0.001}$'' in the second panel from the top, mark the spectral region where OVI associated with the HI BLB is expected, and not seen. The feature labeled ``FPN'' (Fixed Pattern Noise) in the first and second panel from the top, is a known FUSE Lif2A segment instrumental feature (e.g. Sembach et al., 2004)
\label{fig7}}
\end{figure}

\begin{figure}
\plotone{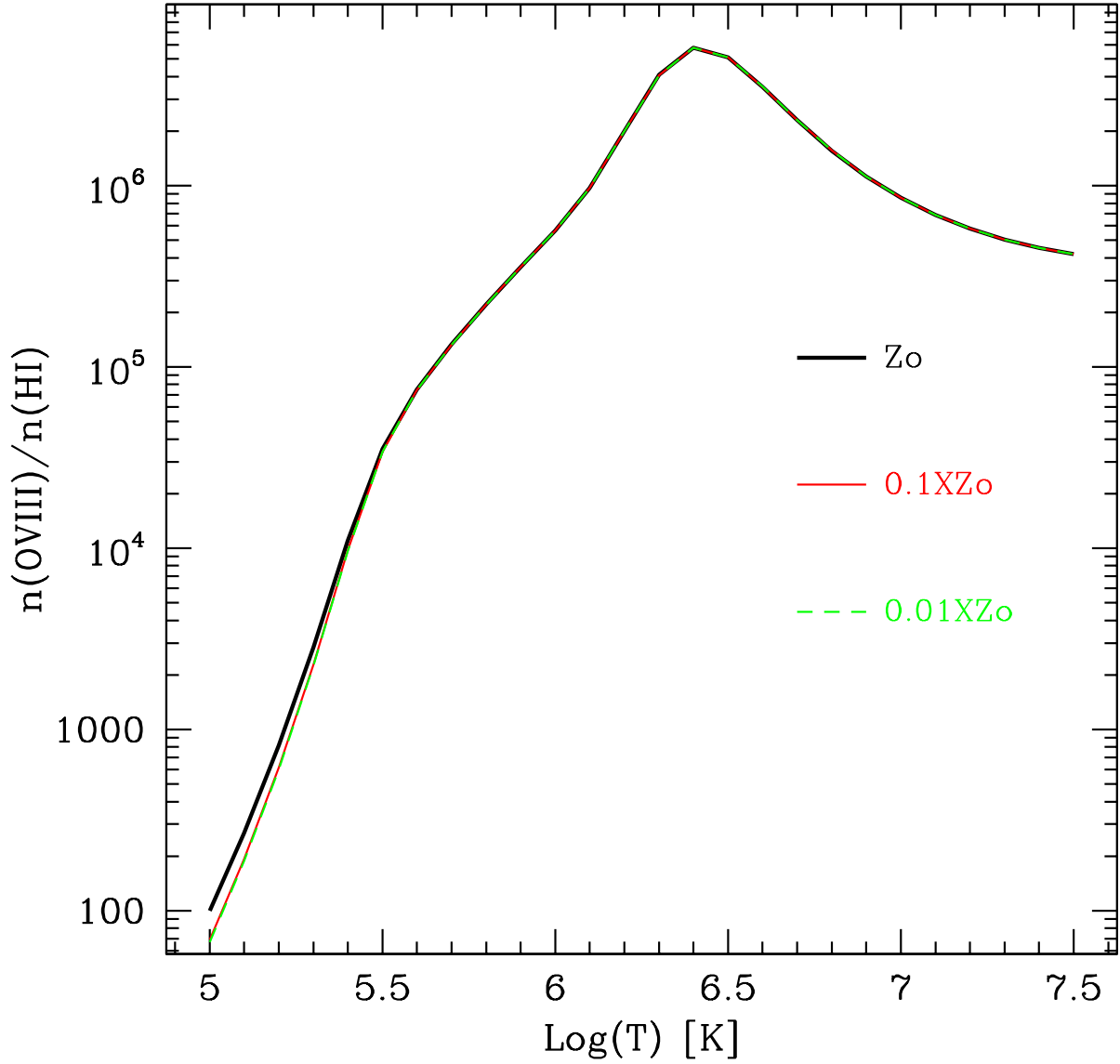}
\caption{OVIII to HI fractional abundance ratio (i.e. ionization correction) as a function of the temperature of the gas, for three different absolute metallicities: $Z=Z_{\odot}$ (black solid curve), $Z=0.1 Z_{\odot}$ (red solid curve) and $Z=0.01 Z_{\odot}$ (green dashed curve). The ionization balance of the gas, at equilibrium, is virtually independent on the value of $Z$, and so is the ionization correction. 
\label{fig7}}
\end{figure}

\end{document}